\documentclass[12pt,a4paper]{amsart}

\usepackage[latin1]{inputenc}     
\usepackage{amsmath,amssymb,amsthm}
\usepackage{latexsym}
\usepackage{amsfonts}
\usepackage{bbm,bm,dsfont} 
\usepackage{color} 
\usepackage{graphicx}

\numberwithin{equation}{section} 

\theoremstyle{definition}
\newtheorem{proposition}{Proposition}

\newtheorem{example}{Example}
\newtheorem{remark}{Remark}

\newcommand{\hA}{\mathcal{A}}
\newcommand{\hB}{\mathcal{B}}

\newcommand{\sfa}{\mathsf{A}}
\newcommand{\sfb}{\mathsf{B}}

\newcommand{\sfe}{\mathsf{E}}
\newcommand{\sff}{\mathsf{F}}
\newcommand{\sfg}{\mathsf{G}}

\newcommand{\sfq}{\mathsf{Q}}

\newcommand{\R}{\mathbb R} 
\newcommand{\Z}{\mathbb Z} 
\newcommand{\C}{\mathbb C} 

\newcommand{\vsigma}{\bm{\sigma}} 
\newcommand{\vb}{\bm{b}} 
\newcommand{\ve}{\bm{e}} 
\newcommand{\vf}{\bm{f}} 

\newcommand{\fii}{\varphi} 
\newcommand{\la}{\lambda} 

\newcommand{\hi}{\mathcal{H}} 

\newcommand{\id}{{\mathds1}} 

\newcommand{\tr}[1]{\mathrm{tr}\left[#1\right]} 
\def\<{\langle} 
\def\>{\rangle} 
\newcommand{\kb}[2]{|#1 \rangle\langle #2|} 

\newcommand{\Eo}{\mathsf{E}} 
\newcommand{\Fo}{\mathsf{F}} 
\newcommand{\Po}{\mathsf{P}} 
\newcommand{\Mo}{\mathsf{M}} 
\newcommand{\No}{\mathsf{N}} 
\newcommand{\G}{\mathsf{G}} 

\def\d{{\mathrm d}} 
\newcommand{\ov}{\overline} 
\newcommand{\CHI}[1]{\ensuremath{ \chi\raisebox{-1ex}{$\scriptstyle #1$} }} 

\newcommand{\m}{\bm m} 
\newcommand{\sgn}{\mathrm{sgn}\,} 

\renewcommand{\leq}{\leqslant}
\renewcommand{\le}{\leqslant}

\renewcommand{\ge}{\geqslant}

\begin{document}

\title[Naimark method for compatibilty]{Naimark dilations of qubit POVMs and joint measurements}

\author{Juha-Pekka Pellonp\"a\"a}
\email{juhpello@utu.fi}
\address{Department of Physics and Astronomy, University of Turku, FI-20014 Turku, Finland}
\author{S\'ebastien Designolle}
\email{sebastien.designolle@unige.ch}
\address{Group of Applied Physics, University of Geneva, 1211 Geneva, Switzerland}
\author{Roope Uola}
\email{roope.uola@unige.ch}
\address{Group of Applied Physics, University of Geneva, 1211 Geneva, Switzerland}

\begin{abstract}
Measurement incompatibility is one of the cornerstones of quantum theory. This phenomenon appears in many forms, of which the concept of non-joint measurability has received considerable attention in the recent years. In order to characterise this non-classical phenomenon, various analytical and numerical methods have been developed. The analytical approaches have mostly concentrated on the qubit case, as well as to scenarios involving sets of measurements with symmetries, such as position and momentum or sets of mutually unbiased bases. The numerical methods can, in principle, decide any finite-dimensional and discrete joint measurability problem, but they naturally have practical limitations in terms of computational power. These methods exclusively start from a given set of measurements and ask whether the set possesses incompatibility. Here, we take a complementary approach by asking which measurements are compatible with a given measurement. It turns out, that this question can be answered in full generality through a minimal Naimark dilation of the given measurement: the set of interest is exactly those measurements that have a block-diagonal representation in such dilation. We demonstrate the use of the technique through various qubit examples, leading to an alternative characterisation of all compatible pairs of binary qubit measurements, which retrieves the celebrated Busch criterion. We further apply the technique to special examples of trinary and continuous qubit measurements.
\end{abstract}

\maketitle


\section{Introduction}

The act of measurement is in the core of any physical theory. In quantum theory, this process provides a window between the microscopic and macroscopic worlds. It is through this fundamental action that one collects new data, verifies and falsifies predictions, and ultimately develops novel theoretical models. Given its fundamental position and strong mathematical grounds, the theory of quantum measurements keeps naturally lending itself to open questions, such as the measurement problem \cite{busch16}, uncertainty relations \cite{buschrmp2014,coles17}, and practical applications of quantum information theory \cite{Gisin2002Review,Degen2017,RandomnessReview}. In this manuscript, we contribute to a specific open question of characterising a central quantum-to-classical border within quantum measurement theory, that is, the threshold between sets of measurements that allow and do not allow a simultaneous readout.

In quantum theory, the traditional description for the act of measurement is given through Hermitian operators, also called observables. By now, it is self-evident that such presentation differs from classical physics, in that whereas all classical measurements can be performed simultaneously, this is no longer true in the quantum scenario. Such discrepancy sets fundamental limitations on the possibility of coding information about non-commuting quantities into a quantum state, as shown by the famous preparation uncertainty relations of Heisenberg and Robertson.

By now, the quantum information theoretic representation of quantum measurements has deviated from the notion of observables due to the introduction of more general positive operator-valued measures (POVMs for short). POVMs offer various advantages over their predecessors in that they, e.g., better capture realistic measurement implementations \cite{busch16,Guryanova20}, can perform better in discriminating quantum states \cite{oszmaniec19,uola19b}, and offer various fine-tuned notions of measurement incompatibility \cite{Pekka03,heinosaari10,heinosaari16a,oszmaniec17,Ioannou22,Cope22}.

For our purposes, a central property of POVMs is that of joint measurability. This is a generalisation of the notion of commutativity of observables. In short, joint measurability asks whether the measurement data of a given set of measurements can be classically post-processed from the data of a single measurement. On the conceptual level, joint measurability has found various applications in, e.g., quantum correlations \cite{wolf09,quintino14,uola14,uola15,kiukas17}, contextuality \cite{tavakoli19}, quantum state discrimination \cite{skrzypczyk19,carmeli19a,uola19b,oszmaniec19}, quantum communication \cite{guerini19}, and quantum thermodynamics \cite{Beyer22}. Hence, the task of characterising sets of measurements that allow a simultaneous readout has become an important and actively investigated problem not only from the foundational, but also from the practical perspective \cite{Zhou16,Designolle21,Hammad2020,Smirne2022}, see also \cite{JMreview} for a recent review. 

Typically, one is interested in characterising the sets of measurements that allow a joint measurement. Here, we take a slightly complementary approach by using a technique that characterises those measurements that are jointly measurable with a given POVM. The technique is based on a minimal Naimark dilation, i.e., representing POVMs as observables in a larger space, and it connects to the sole notion of incompatibility possessed by observables, i.e., non-commutativity. Namely, the set of POVMs that are jointly measurable with a given POVM turns out to be exactly the set of those measurements that have a block-diagonal representation in a minimal dilation of the given POVM. Such block-diagonal measurements are characterised by the commutant of the relevant observable in the minimal dilation space. We want to stress out that this technique has appeared in the past in conceptual works including some of the authors \cite{pellonpaa14,haapasalo15,Pello7} and that connections between joint measurability and commutativity in some dilation space have been reported independently in \cite{JMNaimark,pellonpaa14,Beneduci14,Mitra20}. However, the works \cite{haapasalo15,Pello7} did not concentrate on explicitly characterising joint measurability of given sets of POVMs and the works \cite{JMNaimark,Beneduci14,Mitra20} did not specify a single generally applicable Naimark dilation. Some of the involved dilations even require one to know some joint measurement before the construction of the dilation, i.e.\ one gets the dilation only after one has solved the problem of joint measurability. Here, we stress that one can simply use a minimal dilation of the involved measurements in order to solve the problem of joint measurability. We further note that this dilation is fully constructive. The investigation of joint measurability criteria arising from this process is the main contribution of this manuscript.

Although the minimal dilation technique applies even to infinite-dimensional systems, cf. Ref.~\cite{pellonpaa14}, we concentrate on the qubit case for simplicity. In this setting, the technique provides an alternative way of obtaining the celebrated Busch criterion for joint measurability of two unbiased qubit measurements \cite{busch86} and gives a full characterisation of those qubit effects that are (pairwisely) compatible with a given qubit effect. Moreover, we compare the technique to that of Ref.~\cite{jae19} in the case of two symmetric three-valued qubit POVMs, and provide examples of compatible pairs and triplets of continuous qubit POVMs.

\section{Joint measurability}

We describe quantum measurements as POVMs. In the case of a discrete (finite dimensional) measurement, these are essentially sets of positive semi-definite matrices $\Eo=(E_1,E_2,\ldots,E_N)$ that sum to the identity operator, i.e., $\sum_{i=1}^N E_i=\id$. In the continuous case, a POVM is a normalised (weakly) $\sigma$-additive map from a $\sigma$-algebra $\hA\subseteq 2^\Omega$ to the set of positive operators of a Hilbert space. A discrete POVM $\sfb=(B_1,B_2,\ldots,B_M)$ is called jointly measurable with $\Eo$, if there exists a third POVM ${\sf N}=(N_{ij})$ with $i\in{1,...,N}$ and $j\in{1,...,M}$ such that
\begin{align}
    E_i=\sum_{j=1}^M N_{ij}\label{Eq:JMdef1}\\
    B_j=\sum_{i=1}^N N_{ij}\label{Eq:JMdef2}
\end{align}
for all $i$ and $j$. Otherwise, the POVMs $\Eo$ and $\sfb$ are called incompatible. The POVM ${\sf N}$ is called a joint or parent POVM. This definition generalises directly to continuous POVMs by replacing the sums in Eq.~(\ref{Eq:JMdef1}) and Eq.~(\ref{Eq:JMdef2}) with (total) sets as well as the singletons $i$ and $j$ with measurable sets, i.e., two POVMs $\sfe$ and $\sfb$ (defined on $\sigma$-algebras $\hA\subseteq 2^\Omega$ and $\hB\subseteq 2^\Xi$) are jointly measurable if there is a POVM $\No$ defined on the product $\sigma$-algebra $\hA\otimes\hB\subseteq 2^{\Omega\times\Xi}$ such that $\sfe(X)=\No(X\times\Xi)$ for all $X\in\hA$ and $\sfb(Y)=\No(\Omega\times Y)$ for all $Y\in\hB$.

For basic examples of jointly measurable POVMs, one can choose $\Eo$ and $\sfb$ commuting, i.e., $[E_i,B_j]=0$ for all $i$ and $j$, in which case $N_{ij}=E_iB_j=\sqrt{B_j}E_i\sqrt{B_j}$ is clearly a joint POVM. For a non-commuting example one can take noisy spin measurements defined by
\begin{align*}
    E_{\pm1}=\frac{1}{2}\Big(\id\pm\frac{1}{\sqrt 2}\sigma_1\Big),\\
    B_{\pm1}=\frac{1}{2}\Big(\id\pm\frac{1}{\sqrt 2}\sigma_3\Big).
\end{align*}
In this case, a joint POVM is given by
\begin{align*}
    N_{ij}=\frac{1}{4}\Big[\id\pm\frac{1}{\sqrt 2}(i\sigma_1+j\sigma_3)\Big],\qquad i,j=\pm1.
\end{align*}
One can further show \cite{busch86} that this joint POVM is optimal in the sense that if one increases the length of the Bloch vectors of $E_{\pm1}$ or $B_{\pm1}$, the measurements become incompatible.

In the above examples, the (optimal) joint POVM is rather simple to find. Some techniques for finding joint POVMs for more complex scenarios have been reported in the literature based on a so-called adaptive strategy \cite{Uola16} and other ans\"atze \cite{Heinosaari13b,designolle19b,Kiukas22}. Such techniques typically do not use auxiliary systems. In the following, we map the problem of finding joint measurements into a problem of characterising a commutant in a minimal Naimark dilation space. This provides further natural ans\"atze for joint measurements. We focus our attention to the qubit case and demonstrate the technique by building optimal and suboptimal joint measurements for various scenarios.

\section{Naimark dilations of discrete qubit POVMs}\label{Naimark Section}

Let $\hi$ be a two-dimensional (qubit) Hilbert space. By fixing an orthonormal basis $\{\fii_1,\fii_2\}\subset\hi$ we may identify $\hi$ with $\C^2$ via unitary operator $U:\hi\to\C^2$, $U:=\kb{(1,0)}{\fii_1}+\kb{(0,1)}{\fii_2}$,
where the vectors $(1,0)$ and $(0,1)$ constitute the standard basis of the Hilbert space $\C^2$.
In what follows, we identify any operator $O:\,\hi\to\hi$ with the operator ($2\times2$--matrix) $UOU^*$ on $\C^2$ and study only matrices. Clearly, each $2\times2$-matrix $M$ corresponds to a unique operator $U^*MU$ on $\hi$. 

We say that a positive semidefinite $2\times2$--matrix $E$ is an {\it effect} if $\id-E$ is also positive semidefinite; here 
$\id$ is the identity matrix. Any effect $E$ can be written in the form
$$
E=\frac12\sum_{\mu=0}^3 e^\mu\sigma_\mu= \frac12(e^0\id+\ve\cdot\vsigma)
=\frac12\begin{pmatrix}
e^0+e^3 & e^1-i e^2 \\
e^1+i e^2 & e^0-e^3
\end{pmatrix}
$$
where $(e^0,e^1,e^2,e^3)=(e^0,\ve)\in \R^4$, $\|\ve\|:=\sqrt{(e^1)^2+(e^2)^2+(e^3)^2}\le\min\{e^0,2-e^0\}$, and
$$
\sigma_0=\id=\left(\begin{matrix} 1&0\\ 0&1 \end{matrix}\right),\quad
\sigma_1=\left(\begin{matrix} 0&1\\ 1&0 \end{matrix}\right),\quad
\sigma_2=\left(\begin{matrix} 0&-i\\ i&0 \end{matrix}\right),\quad\text{and}\quad
\sigma_3=\left(\begin{matrix} 1&0\\ 0&-1 \end{matrix}\right)
$$
are the Pauli matrices \cite[Chapter 14]{busch16}. In particular, $e^\mu=\tr{E\sigma_\mu}$, $\mu=0,1,2,3$, $e^0\in[0,2]$, and $|e^j|\le\|\ve\|\le1$, $j=1,2,3$. The eigenvalues of $E$ are $\frac{1}{2}(e^0\pm\|\ve\|)\in[0,1]$ so that $E$ is of rank 1 if and only if $e^0=\|\ve\|\ne0$. Especially, $E$ is a rank-1 (resp.\ rank-2) projection exactly when $e^0=\|\ve\|=1$ (resp.\ $e^0=2$ and $\|\ve\|=0$).

Define then the parameters
\begin{eqnarray*}
c^\pm(E) &:=& \sqrt{\frac{(e^0\pm\|\ve\|)(\|\ve\|\pm e^3)}{4\|\ve\|}}\ge 0,\\
d^\pm(E) &:=& \pm\big(e^1-i e^2\big)
\sqrt{\frac{e^0\pm\|\ve\|}{4\|\ve\|(\|\ve\|\pm e^3)}}\in\C,
\end{eqnarray*}
if $\|\ve\|\ne | e^3|$, and
\begin{eqnarray}\label{yht}
c^\pm(E) &:=&\frac{1\pm\sgn e^3}2\sqrt{\frac{e^0\pm|e^3|}2}\ge 0,\\
d^\pm(E) &:=& \frac{1\mp\sgn e^3}2\sqrt{\frac{e^0\pm|e^3|}2}\ge 0,\nonumber
\end{eqnarray}
if $\|\ve\|= | e^3|$ (i.e., $e^1=e^2=0$); here $\sgn x:=1$ when $x\ge 0$ and $-1$ otherwise.
 Clearly, $\overline{c^+(E)}c^-(E)+\overline{d^+(E)}d^-(E)=0$. 
We can write the spectral decomposition $E=E^++E^-$ where
$$
E^\pm:=
\begin{pmatrix}
|c^\pm(E)|^2 & \overline{c^\pm(E)}d^\pm(E) \\
c^\pm(E)\overline{d^\pm(E)} & |d^\pm(E)|^2 
\end{pmatrix}
=
\frac{e^0\pm\|\ve\|}{2}\cdot\frac1{2\|\ve\|}
\begin{pmatrix}
\|\ve\|\pm e^3 & \pm(e^1-i e^2) \\
\pm(e^1+i e^2) & \|\ve\|\mp e^3
\end{pmatrix}
$$
are rank-1 or zero effects. If $\|\ve\|=0$ then $E=(e^0/2)\id$ and one has $E^\pm=(e^0/4)(\id\pm\sigma_3)$ by Eq.\ \eqref{yht}; only in this degenerate case the spectral projections are not unique. Note that $E\ne 0$ is of rank 1 if and only if $E^-=0$ (i.e., $E=E^+$ or $c^-(E)=d^-(E)=0$).
Finally, if $M$ is a positive semidefinite rank-1 matrix then $M$ can be written in the form
$$
M=
\begin{pmatrix}
|c|^2 & \overline{c}d \\
c\overline{d} & |d|^2 
\end{pmatrix}
$$
where the complex numbers $c$ and $d$ are unique if we assume that either $c> 0$ and $d\in\C$ or $c=0$ and $d>0$ (that is, $(c,d)\in\big(\R_+\times\C\big)\cup\big(\{0\}\times\R_+\big)$ where $\R_+$ is the set of positive reals).

Let $\Eo=(E_1,E_2,\ldots,E_N)$ be an {\it $N$-valued POVM} of $\C^2$, i.e., the {\it non-zero} effects 
$$
E_i=
\frac12\sum_{\mu=0}^3 e_i^\mu\sigma_\mu= \frac12(e_i^0\id+\ve_i\cdot\vsigma),\qquad i=1,2,\ldots N,
$$ 
sum to the identity matrix $\id$. Write, as above, $E_i=E_i^++E_i^-$ where
$$
E_i^\pm:=
\begin{pmatrix}
|c^\pm(E_i)|^2 & \overline{c^\pm(E_i)}d^\pm(E_i) \\
c^\pm(E_i)\overline{d^\pm(E_i)} & |d^\pm(E_i)|^2
\end{pmatrix}.
$$
Let $m_i\in\{1,2\}$ be the rank of $E_i$ and form the multiplicity or rank vector of $\Eo$,
$$
\m:=(m_1,m_2,\ldots m_N)
$$
whose $\ell^1$-norm is $\|\m\|_1=\sum_{i=1}^N m_i$.
If $E_i$ is rank-1 ($m_i=1$) define ${\bm c}(E_i):=c^+(E_i)$ and
${\bm d}(E_i):=d^+(E_i)$.
If $E_i$ is of rank 2 ($m_i=2$) define ${\bm c}(E_i):=\big(c^+(E_i),c^-(E_i)\big)$ and
${\bm d}(E_i):=\big(d^+(E_i),d^-(E_i)\big)$ which satisfy the orthogonality relation 
$\overline{c^+(E_i)}c^-(E_i)+\overline{d^+(E_i)}d^-(E_i)=0$.

Now the vectors
$$
{\bm c}:=\big({\bm c}(E_1),{\bm c}(E_2),\ldots,{\bm c}(E_N)\big),\qquad
{\bm d}:=\big({\bm d}(E_1),{\bm d}(E_2),\ldots,{\bm d}(E_N)\big)
$$
belong to the minimal Naimark dilation space $\C^{\|\m\|_1}$. Indeed, define an isometry $J_{\bm c,\bm d}:\,\C^2\to\C^{\|\m\|_1}$ via
$$
J_{\bm c,\bm d}:=\kb{{\bm c}}{(1,0)}+\kb{{\bm d}}{(0,1)}=
\begin{pmatrix}
c_1 & d_1  \\
c_2 & d_2  \\
\vdots & \vdots  \\
c_{\|\m\|_1} & d_{\|\m\|_1}  \\
\end{pmatrix}
$$ 
where we have denoted briefly ${\bm c}=(c_1,\ldots,c_{\|\m\|_1})$ and ${\bm d}=(d_1,\ldots,d_{\|\m\|_1})$.
Especially, ${\bm c}$ and ${\bm d}$ are orthonormal vectors (i.e., $\|{\bm c}\|=\|{\bm d}\|=1$ and $\<{\bm c}|{\bm d}\>=0$),
 $J_{\bm c,\bm d}^*J_{\bm c,\bm d}=\id$, and $J_{\bm c,\bm d}J_{\bm c,\bm d}^*=\kb{\bm c}{\bm c}+\kb{\bm d}{\bm d}$ is a projection on $\C^{\|\m\|_1}$. 
 In addition,
 $$
 (c_i,d_i)\in\big(\R_+\times\C\big)\cup\big(\{0\}\times\R_+\big)
 $$
for all $i=1,2,\ldots,{\|\m\|_1}$.
Let $\{b_k\}_{k=1}^{\|\m\|_1}$ be the standard (orthonormal) basis of $\C^{\|\m\|_1}$. Define $K_0:=0$,
$K_i:=\sum_{k=1}^i m_k$, $i\in\{1,\ldots,N\}$,
and projections
$$
P_i:=\sum_{k=1+K_{i-1}}^{K_i}\kb{b_k}{b_k}
$$
so that $\Po=(P_1,\ldots,P_N)$ is a projection valued measure (PVM).
Since $E_i=J_{\bm c,\bm d}^*P_i J_{\bm c,\bm d}$ the triple $(\C^{\|\m\|_1},\Po,J_{\bm c,\bm d})$ is a {\it minimal\footnote{Clearly, the vectors $P_i\bm c$ and $P_i\bm d$, $i=1,\ldots,N$, span $\C^{\|\m\|_1}$.} Naimark dilation} of $\Eo$, see, e.g., Ref.~\cite{busch16}.

It should be stressed that any orthonormal vectors $\bm c,$ $\bm d\in\C^{\|\m\|_1}$ can be used to define an isometry $J_{\bm c,\bm d}:=\kb{{\bm c}}{(1,0)}+\kb{{\bm d}}{(0,1)}$ and POVM $\sfe$ via $E_i:=J_{\bm c,\bm d}^*P_i J_{\bm c,\bm d}$ but $\m$ is not necessarily the multiplicity vector of $\sfe$. It may happen that some $m_i=2$ but the rank of $E_i$ is 0 or 1.
However, if $\sfe$ is a rank-1 POVM (i.e., $m_i\equiv1$ so $\|\m\|_1=N$) then one has the following uniqueness result.

\begin{proposition}
Let $N$ be a positive integer.
Then there is a bijection between the set of $N$-valued rank-1 POVMs and 
the set of orthonormal vectors $\bm c,\,\bm d\in\C^{N}$ such that 
 $(c_i,d_i)\in\big(\R_+\times\C\big)\cup\big(\{0\}\times\R_+\big)$
for all $i=1,2,\ldots,N$.
\end{proposition}

\begin{example}
In this example, we characterise all $N$-valued qubit POVMs whose multiplicity vectors $\m$ have the same length $\|\m\|_1=4$ using the above dilation technique.
We have the following nontrivial cases $\m=(1,1,1,1)$ [$N=4$], $\m=(2,1,1)$, $\m=(1,2,1)$, $\m=(1,1,2)$ [$N=3$], and $\m=(2,2)$ [$N=2$]. In all these cases, the dilation space is $\C^4$ and the isometry
$$
J_{\bm c,\bm d}=
\begin{pmatrix}
c_1 & d_1  \\
c_2 & d_2  \\
c_3 & d_3 \\
c_4 & d_4  \\
\end{pmatrix}
$$
where 
$
 (c_i,d_i)\in\big(\R_+\times\C\big)\cup\big(\{0\}\times\R_+\big)
$,
$i=1,2,3,4$, are such that $\|{\bm c}\|=\|{\bm d}\|=1$ and $\<{\bm c}|{\bm d}\>=0$.
By varying $\bm c$ and $\bm d$ we get all POVMs with $\|\m\|_1=4$.
The above cases differ on the definition of the projections $P_i$:

\begin{itemize}
\item $\boxed{\m=(1,1,1,1)}$ Now $P_i=\kb{b_i}{b_i}$ for all $i$, e.g.,
$$
P_3=
\kb{b_3}{b_3}=
\begin{pmatrix}
0 \\
0 \\
1 \\
0 \\
\end{pmatrix}
\begin{pmatrix}
0 & 0 & 1 & 0 
\end{pmatrix}
=
\begin{pmatrix}
0 & 0 & 0 & 0 \\
0 & 0 & 0 & 0 \\
0 & 0 & 1 & 0 \\
0 & 0 & 0 & 0 \\
\end{pmatrix}
$$
and
$$
 E_3=J_{\bm c,\bm d}^*P_3J_{\bm c,\bm d}=
\begin{pmatrix}
\ov c_1 & \ov c_2 & \ov c_3 & \ov c_4 \\
\ov d_1 & \ov d_2 & \ov d_3 & \ov d_4 \\
\end{pmatrix}
\begin{pmatrix}
0 & 0 & 0 & 0 \\
0 & 0 & 0 & 0 \\
0 & 0 & 1 & 0 \\
0 & 0 & 0 & 0 \\
\end{pmatrix}
\begin{pmatrix}
c_1 & d_1  \\
c_2 & d_2  \\
c_3& d_3  \\
c_4 & d_4  \\
\end{pmatrix}
=\begin{pmatrix}
|c_3|^2 & \overline c_3 d_3  \\
c_3\overline d_3 & |d_3|^2  \\
\end{pmatrix}.
$$
Similarly,
$$
E_i=
\begin{pmatrix}
|c_i|^2 & \overline c_i d_i  \\
c_i\overline d_i & |d_i|^2  \\
\end{pmatrix}
$$
for all $i=1,2,3,4$, and one can check that
$$
\sum_{i=1}^4E_i=
\sum_{i=1}^4\begin{pmatrix}
|c_i|^2 & \overline c_i d_i  \\
\overline d_i c_i & |d_i|^2  \\
\end{pmatrix}
=
\begin{pmatrix}
\|\bm c\|^2 & \<\bm c|\bm d\>  \\
\<\bm d|\bm c\> & \|\bm d\|^2  \\
\end{pmatrix}
=
\begin{pmatrix}
1 & 0  \\
0 & 1 \\
\end{pmatrix}=\id.
$$

\item $\boxed{\m=(2,1,1)}$ The cases
$\m=(2,1,1)$, $\m=(1,2,1)$, and $\m=(1,1,2)$ are essentially the same, so we study only the first one.
Now $P_1=\kb{b_1}{b_1}+\kb{b_2}{b_2}$, $P_2=\kb{b_3}{b_3}$, and $P_3=\kb{b_4}{b_4}$ showing that
$$
E_1=
\begin{pmatrix}
|c_1|^2 & \overline c_1 d_1  \\
c_1\overline d_1 & |d_1|^2  \\
\end{pmatrix}
+
\begin{pmatrix}
|c_2|^2 & \overline c_2 d_2  \\
c_2\overline d_2 & |d_2|^2  \\
\end{pmatrix},\quad
E_2=
\begin{pmatrix}
|c_3|^2 & \overline c_3 d_3  \\
c_3\overline d_3 & |d_3|^2  \\
\end{pmatrix},\quad
E_3=
\begin{pmatrix}
|c_4|^2 & \overline c_4 d_4  \\
c_4\overline d_4 & |d_4|^2  \\
\end{pmatrix}.
$$
Note that $E_1$ is of rank 2 exactly when $c_1d_2\ne c_2d_1$ (which we must assume). 

\item $\boxed{\m=(2,2)}$ Now $P_1=\kb{b_1}{b_1}+\kb{b_2}{b_2}$, $P_2=\kb{b_3}{b_3}+\kb{b_4}{b_4}$ so that
$$
E_1=
\begin{pmatrix}
|c_1|^2 & \overline c_1 d_1  \\
c_1\overline d_1 & |d_1|^2  \\
\end{pmatrix}
+
\begin{pmatrix}
|c_2|^2 & \overline c_2 d_2  \\
c_2\overline d_2 & |d_2|^2  \\
\end{pmatrix},\quad
E_2=
\begin{pmatrix}
|c_3|^2 & \overline c_3 d_3  \\
c_3\overline d_3 & |d_3|^2  \\
\end{pmatrix}
+
\begin{pmatrix}
|c_4|^2 & \overline c_4 d_4  \\
c_4\overline d_4 & |d_4|^2  \\
\end{pmatrix}.
$$
Further assumptions $c_1d_2\ne c_2d_1$ and $c_3d_4\ne c_4d_3$ yield rank-2 effects $E_1$ and $E_2$.
\end{itemize}
\end{example}

\begin{example}In this example, we study 2-valued qubit POVMs.
It is easy to see that we have (essentially) the following cases:

\begin{itemize}
\item $\boxed{\m=(1,1)}$ Now
$$
E_1=
\begin{pmatrix}
|c_1|^2 & \overline c_1 d_1  \\
c_1\overline d_1 & |d_1|^2  \\
\end{pmatrix},
\quad
E_2=
\begin{pmatrix}
|c_2|^2 & \overline c_2 d_2  \\
c_2\overline d_2 & |d_2|^2  \\
\end{pmatrix}
=\id-E_1=\begin{pmatrix}
1-|c_1|^2 & -\overline c_1 d_1  \\
-c_1\overline d_1 & 1-|d_1|^2  \\
\end{pmatrix}
$$
where
$
 (c_i,d_i)\in\big(\R_+\times\C\big)\cup\big(\{0\}\times\R_+\big)
$,
$i=1,2$, are such that $\|{\bm c}\|=\|{\bm d}\|=1$ and $\<{\bm c}|{\bm d}\>=0$.
\item $\boxed{\m=(2,1)}$ Now
$$
E_1=
\begin{pmatrix}
|c_1|^2 & \overline c_1 d_1  \\
c_1\overline d_1 & |d_1|^2  \\
\end{pmatrix}
+
\begin{pmatrix}
|c_2|^2 & \overline c_2 d_2  \\
c_2\overline d_2 & |d_2|^2  \\
\end{pmatrix},\quad
E_2=
\begin{pmatrix}
|c_3|^2 & \overline c_3 d_3  \\
c_3\overline d_3 & |d_3|^2  \\
\end{pmatrix}
$$
where 
$
 (c_i,d_i)\in\big(\R_+\times\C\big)\cup\big(\{0\}\times\R_+\big)
$,
$i=1,2,3$, are such that $\|{\bm c}\|=\|{\bm d}\|=1$, $\<{\bm c}|{\bm d}\>=0$, and $c_1d_2\ne c_2d_1$.

\item $\boxed{\m=(2,2)}$ See the preceding example.
\end{itemize}

\end{example}

\begin{remark}
It is well known that, if $\Eo$ is extremal in the convex set of all POVMs, then $N\le4$. Hence, if $N>5$ then $\Eo$ can be written as a barycentre of extremal POVMs (by adding zero effects if necessary) \cite{Chiribella10}. For this reason, we usually assume that $1<N\le 4$ (the case $N=1$ is trivial).
Moreover, all these cases can be incorporated into the case $N=4$ by adding zeros: if $N=3$ we set $E_4=0$ and write $\Eo=(E_1,E_2,E_3,0)$ and, if $N=2$, $\Eo=(E_1,E_2,0,0)$.
Now we may also add zero components to the (minimal) vectors $\bm c,\,\bm d$.
\end{remark}

\section{Jointly measurable discrete qubit POVMs}

Let $\Eo=(E_1,E_2,\ldots,E_N)$ be an $N$-valued qubit POVM with the multiplicity vector $\m$ and the minimal Naimark dilation $(\C^{\|\m\|_1},\Po,J_{\bm c,\bm d})$ as before. To characterise {\it all} qubit POVMs $\sfb$ jointly measurable with $\Eo$, one can pick any POVM $\Fo=(F_1,F_2,\ldots,F_M)$ of the dilation space $\C^{\|\m\|_1}$ such that each effect
$F_j$ is decomposable\footnote{That is, $\Fo$ and $\Po$ commute: $[F_j,P_i]= 0$.} with respect to $\Po$, that is,
$$
F_j=\bigoplus_{i=1}^N F_{ij}
$$
where $\Fo_i=(F_{i1},F_{i2},\ldots,F_{iM})$ is a POVM of $\C^{m_{i}}$ (whose identity operator is $P_i$). Now it may happen that an effect $F_{ij}$ is zero. In the case $m_i=1$ the POVM $\Fo_i$ `is' just a sequence of numbers $f_{ij}\ge 0$ (i.e., $F_{ij}=f_{ij}P_i$) such that  and 
$\sum_{j=1}^M f_{ij}=1$, whereas in the case $m_i=2$, $\Fo_i$ {\it is a qubit POVM.} 
The jointly measurable POVM $\sfb=(B_1,B_2,\ldots,B_M)$  is of the form
$$
B_j=J_{\bm c,\bm d}^* F_j J_{\bm c,\bm d}=\sum_{i=1}^NJ_{\bm c,\bm d}^* F_{ij} J_{\bm c,\bm d}
$$
and the joint POVM is ${\sf N}=(N_{ij})$ where
$$
N_{ij}=J_{\bm c,\bm d}^* F_{ij} J_{\bm c,\bm d}.
$$
Indeed, since $\sum_{j=1}^M F_{ij}=P_i$ one sees 
that $\sum_{j=1}^M N_{ij}=J_{\bm c,\bm d}^*P_i J_{\bm c,\bm d}=E_i$ and $\sum_{i=1}^N N_{ij}=B_j$.
It can be shown that we get all compatible POVMs $\sfb$ by using this {\it Naimark dilation technique} \cite{pellonpaa14}. This follows easily since any effect $N_{ij}$ of a joint POVM is majorised by $E_i=(P_iJ_{\bm c,\bm d})^*P_i J_{\bm c,\bm d}$ so that there exists an effect $F_{ij}\le P_i$ for which $N_{ij}=J_{\bm c,\bm d}^* F_{ij} J_{\bm c,\bm d}$ holds.
Next we study the structure of the effects $N_{ij}$.

If $m_i=1$ then $N_{ij}=f_{ij}E_i$.
If $m_i=2$ then we can identify the qubit effect $F_{ij}$ with the matrix
$$
\frac12\sum_{\mu=0}^3 f_{ij}^\mu\sigma_\mu= \frac12(f_{ij}^0\id+\vf_{ij}\cdot\vsigma)
$$
where $(f_{ij}^0,f_{ij}^1,f_{ij}^2,f_{ij}^3)=(f_{ij}^0,\vf_{ij})\in \R^4$ and $\|\vf_{ij}\|\le\min\{f_{ij}^0,2-f_{ij}^0\}$. Indeed,
since now $P_i=\sum_{k=K_i-1}^{K_i}\kb{b_k}{b_k}$,
$$
F_{ij}=Y_i^*\ \frac12\sum_{\mu=0}^3 f_{ij}^\mu\sigma_\mu\  Y_i=\frac12\sum_{\mu=0}^3 f_{ij}^\mu Y_i^*\sigma_\mu Y_i
$$
where $Y_i=\kb{(1,0)}{b_{K_i-1}}+\kb{(0,1)}{b_{K_i}}$ is an isometry. 
From the equations
$$
J_{\bm c,\bm d}^*Y_i^*\sigma_0 Y_iJ_{\bm c,\bm d}=
\begin{pmatrix}
|c_{K_i-1}|^2 & \overline c_{K_i-1} d_{K_i-1}  \\
c_{K_i-1}\overline d_{K_i-1} & |d_{K_i-1}|^2  \\
\end{pmatrix}
+
\begin{pmatrix}
|c_{K_i}|^2 & \overline c_{K_i} d_{K_i}  \\
c_{K_i}\overline d_{K_i} & |d_{K_i}|^2  \\
\end{pmatrix},
$$
$$
J_{\bm c,\bm d}^*Y_i^*\sigma_1 Y_iJ_{\bm c,\bm d}=
\begin{pmatrix}
c_{K_i} \overline c_{K_i-1} + c_{K_i-1} \overline c_{K_i} & d_{K_i} \overline c_{K_i-1} + d_{K_i-1} \overline c_{K_i}   \\
c_{K_i} \overline d_{K_i-1}+c_{K_i-1} \overline d_{K_i} & d_{K_i} \overline d_{K_i-1} + d_{K_i-1} \overline d_{K_i}  \\
\end{pmatrix},
$$
$$
J_{\bm c,\bm d}^*Y_i^*\sigma_2 Y_iJ_{\bm c,\bm d}=
i\begin{pmatrix}
-c_{K_i} \overline c_{K_i-1} + c_{K_i-1} \overline c_{K_i} & -d_{K_i} \overline c_{K_i-1} + d_{K_i-1} \overline c_{K_i}   \\
-c_{K_i} \overline d_{K_i-1}+c_{K_i-1} \overline d_{K_i} & -d_{K_i} \overline d_{K_i-1} + d_{K_i-1} \overline d_{K_i}  \\
\end{pmatrix},
$$
$$
J_{\bm c,\bm d}^*Y_i^*\sigma_3 Y_iJ_{\bm c,\bm d}=
\begin{pmatrix}
|c_{K_i-1}|^2 & \overline c_{K_i-1} d_{K_i-1}  \\
c_{K_i-1}\overline d_{K_i-1} & |d_{K_i-1}|^2  \\
\end{pmatrix}
-
\begin{pmatrix}
|c_{K_i}|^2 & \overline c_{K_i} d_{K_i}  \\
c_{K_i}\overline d_{K_i} & |d_{K_i}|^2  \\
\end{pmatrix}
$$
one can calculate
$$
N_{ij}=J_{\bm c,\bm d}^* F_{ij} J_{\bm c,\bm d}=
\frac12\sum_{\mu=0}^3 f_{ij}^\mu \cdot 
J_{\bm c,\bm d}^* Y_i^* \sigma_\mu Y_i J_{\bm c,\bm d}.
$$
Next we give some examples.

\subsection{Compatible effects}

In this section we take advantage of the method presented above to derive the criterion on joint measurability of two two-valued (unbiased) qubit POVMs first presented by Paul Busch in 1986 \cite{busch86}.

Fix a two-valued qubit POVMs $\Eo=(E_1,E_2)$, $E_1+E_2=\id$, and all related notions as in Section \ref{Naimark Section}. Note that $\Eo$ is fully determined by the (nontrivial) effect $E_1$ which we denote briefly by $E$. Next we characterise all two-valued qubit POVMs $\sfb=(B,\id-B)$ (i.e., effects $B$) which are jointly measurable with $\Eo$. 
 We have (essentially) three cases:
\begin{itemize}
\item $\boxed{\m=(1,1)}$  
Let $e_1,\,e_2\in[0,1]$ be arbitrary and 
$$
B=J_{\bm c,\bm d}^*\begin{pmatrix}
e_1 & 0  \\
0 & e_2  \\
\end{pmatrix}J_{\bm c,\bm d}=e_1E_1+e_2E_2=e_2\id+(e_1-e_2)E
$$
showing that the effects $E$ and $B$ commute.

\item $\boxed{\m=(2,1)}$ 
Let $F=\begin{pmatrix}
f_{11} & f_{12}  \\
f_{21} & f_{22}  \\
\end{pmatrix}$ be any effect and $e_3\in[0,1]$. Then
$$
B=J_{\bm c,\bm d}^*\begin{pmatrix}
f_{11} & f_{12}  & 0 \\
f_{21} & f_{22} & 0  \\
0 & 0 & e_3  \\
\end{pmatrix}J_{\bm c,\bm d}=
\begin{pmatrix}
\ov c_1 & \ov c_2 & \ov c_3  \\
\ov d_1 & \ov d_2 & \ov d_3  \\
\end{pmatrix}
\begin{pmatrix}
f_{11} & f_{12}  & 0 \\
f_{21} & f_{22} & 0  \\
0 & 0 & e_3  \\
\end{pmatrix}
\begin{pmatrix}
c_1 & d_1  \\
c_2 & d_2  \\
c_3& d_3  \\
\end{pmatrix}.
$$

\item $\boxed{\m=(2,2)}$ 

Let $F=\begin{pmatrix}
f_{11} & f_{12}  \\
f_{21} & f_{22}  \\
\end{pmatrix}$
and $G=\begin{pmatrix}
g_{11} & g_{12}  \\
g_{21} & g_{22}  \\
\end{pmatrix}$
 be any effects so that
 $$
 B=J_{\bm c,\bm d}^*\begin{pmatrix}
f_{11} & f_{12}  & 0 & 0\\
f_{21} & f_{22} & 0 & 0 \\
0 & 0 & g_{11} & g_{12}  \\
0 & 0 & g_{21} & g_{22}\end{pmatrix}J_{\bm c,\bm d}=
\begin{pmatrix}
\ov c_1 & \ov c_2 & \ov c_3 & \ov c_4 \\
\ov d_1 & \ov d_2 & \ov d_3 & \ov d_4 \\
\end{pmatrix}
\begin{pmatrix}
f_{11} & f_{12}  & 0 & 0\\
f_{21} & f_{22} & 0 & 0 \\
0 & 0 & g_{11} & g_{12}  \\
0 & 0 & g_{21} & g_{22}\end{pmatrix}\begin{pmatrix}
c_1 & d_1  \\
c_2 & d_2  \\
c_3& d_3  \\
c_4 & d_4  \\
\end{pmatrix}.
$$
\end{itemize}

Consider then an {\it unbiased} two-outcome qubit POVM $(E,\id-E)$.
Up to an irrelevant unitary, one can write
\begin{equation*}
  E=\frac{\id+\ve\cdot\vsigma}{2}=\frac{\id+a\sigma_3}{2},
\end{equation*}
where $\ve=(0,0,a)$ and $|a|\leq1$.
The case $|a|=1$ is trivial so that we restrict to $|a|<1$ in the following.
Therefore we have $\m=(2,2)$ and, by using the general procedure introduced previously, we write
\begin{equation*}
  E=J^*\begin{pmatrix}1&0&0&0\\0&1&0&0\\0&0&0&0\\0&0&0&0\end{pmatrix}J\quad\text{where}\quad J=\begin{pmatrix}\sqrt{\frac{1+a}{2}}&0\\0&\sqrt{\frac{1-a}{2}}\\0&\sqrt{\frac{1+a}{2}}\\\sqrt{\frac{1-a}{2}}&0\end{pmatrix}.
\end{equation*}

Now let us consider another (possibly biased) two-outcome qubit POVM $(B,\id-B)$ jointly measurable with $(E,\id-E)$.
We know that
\begin{equation}
  B=J^*\begin{pmatrix}F&0\\0&G\end{pmatrix}J,
  \label{eqn:bjma}
\end{equation}
where $F$ and $G$ are two qubit effects.
In the Pauli basis, equation \eqref{eqn:bjma} gives
\begin{equation*}
  \left\{\begin{array}{l}b^0=\frac{f^0+g^0}{2}+a\cdot\frac{f^3+g^3}{2}\\b^1=\sqrt{1-a^2}\cdot\frac{f^1+g^1}{2}\\b^2=\sqrt{1-a^2}\cdot\frac{f^2-g^2}{2}\\b^3=\frac{f^3-g^3}{2}+a\cdot\frac{f^0-g^0}{2}\end{array}\right.
\end{equation*}
where $B=(b^0\id+\vb\cdot\vsigma)/2$ and similarly for $F$ and $G$. Define $m(x):=\min\{x,2-x\}$, $x\in[0,2]$. 
Then we can use $(f^1)^2+(f^2)^2+(f^3)^2\leq [m(f^0)]^2$, $(g^1)^2+(g^2)^2+(g^3)^2\leq [m(g^0)]^2$, and the Cauchy--Schwarz inequality to get
\begin{align}
  &\ve^2+\vb^2-(\ve\cdot\vb)^2=a^2+(1-a^2)\left[\left(\frac{f^1+g^1}{2}\right)^2+\left(\frac{f^2-g^2}{2}\right)^2+\left(\frac{f^3-g^3}{2}+a\cdot\frac{f^0-g^0}{2}\right)^2\right]\nonumber\\
  &\leq a^2+(1-a^2)\left[\left(\frac{m(f^0)+m(g^0)}{2}\right)^2+\left(a\cdot\frac{f^0-g^0}{2}\right)^2+2a\cdot\frac{f^3-g^3}{2}\cdot\frac{f^0-g^0}{2}\right]\nonumber\\
  &\leq a^2+(1-a^2)\left[\left(\frac{m(f^0)+m(g^0)}{2}\right)^2+\left(a\cdot\frac{f^0-g^0}{2}\right)^2+2|a|\cdot\frac{m(f^0)+m(g^0)}{2}\cdot\left|\frac{f^0-g^0}{2}\right|\right]\nonumber\\
  &\leq a^2+(1-a^2)\left[\frac{m(f^0)+m(g^0)+|f^0-g^0|}{2}\right]^2\leq1,\nonumber
\end{align}
so that in the end we have the following (equivalent) inequality \cite[Prop.\ 14.1]{busch16}:
\begin{equation*}
  \left\|\ve+\vb\right\|+\left\|\ve-\vb\right\|=\sqrt{\ve^2+\vb^2+2\ve\cdot\vb}+\sqrt{\ve^2+\vb^2-2\ve\cdot\vb}\leq\left|1+\ve\cdot\vb\right|+\left|1-\ve\cdot\vb\right|=2.
\end{equation*}
Thus, we have proven that any qubit effect $B$ compatible with $E$ satisfies the above Busch's criterion \cite{busch86}. Actually, this holds for any compatible pair of (possibly biased) qubit effects  \cite[Prop.\ 14.2]{busch16}.

Conversely, if we take any \emph{unbiased} two-outcome qubit POVM such that $\ve^2+\vb^2\leq1+(\ve\cdot\vb)^2$, the following choice gives rise to valid \emph{positive} $F$ and $G$ satisfying equation \eqref{eqn:bjma}:
\begin{equation*}
  \left\{\begin{array}{l}f^0=g^0=b^0=1\quad\text{(unbiased)}\\f^1=g^1=\frac{b^1}{\sqrt{1-a^2}}\\f^2=-g^2=\frac{b^2}{\sqrt{1-a^2}}\\f^3=-g^3=b^3\end{array}\right. 
\end{equation*}
that is, $E$ and $B$ are compatible. 
Finally, we note that there exist incompatible (biased) effects $E$ and $B$ such that  $\ve^2+\vb^2\leq1+(\ve\cdot\vb)^2$ holds. For instance, take $e^0=\sqrt{15}/4$, $\ve=(0,0,e^0)$, 
$b^0=\frac14$, and $\vb=\frac14(1,0,0)$ to get $\ve^2+\vb^2-(\ve\cdot\vb)^2=1$. If the corresponding (rank-1) effects 
$$
E=\frac{\sqrt{15}}4\begin{pmatrix}1&0\\0&0\end{pmatrix},\qquad
B=\frac18\begin{pmatrix}1&1\\1&1\end{pmatrix}
$$
had a joint POVM, i.e., $E=N_{11}+N_{12}$ and $B=N_{11}+N_{21}$, then $N_{11}\le E$ and $N_{11}\le B$ yield $N_{11}=0$ so that 
$$
N_{22}=\id-E-B=\frac18\begin{pmatrix}7-2\sqrt{15}&-1\\-1&7\end{pmatrix}
$$
is not positive (since $7-2\sqrt{15}\approx-0.7$).

\subsection{Three-valued symmetric POVM}
Our symmetry group is the (additive) cyclic group $\Z_3=\{0,1,2\}$ equipped with the addition modulo 3, e.g., $1+2\equiv 0$ (mod 3).
It operates on itself: any $k\in\Z_3$ corresponds to a permutation (i.e., bijection) $b_k(\ell):=k+\ell$ (mod 3) for all $\ell\in\Z_3$.
Furthermore, $\Z_3$ acts in $\C^2$ via the unitary representation $k\mapsto U_k:=R(2k\pi/3)$ where
$$
R(\theta):=\begin{pmatrix}
\cos\theta & -\sin\theta \\
\sin\theta & \cos\theta
\end{pmatrix}\in\rm SO(2)
$$
is the rotation matrix. Hence, we get the mutually commuting unitaries
$$
U_0=\id,\quad 
U_1=U_2^*=\frac12\begin{pmatrix}
-1 & -\sqrt3 \\
\sqrt3 & -1
\end{pmatrix},\quad
U_2=U_1^*=(U_1)^2=\frac12\begin{pmatrix}
-1 & \sqrt3 \\
-\sqrt3 & -1
\end{pmatrix}.
$$
Let $|+\>=(1,0)$ and $|-\>=(0,1)$ denote the eigenvectors of $\sigma_3$. Define a 3-valued covariant POVM $\sfe$ with effects $E_k:=\frac23U_k\kb++U_k^*$, $k\in\Z_3$, and its noisy version $\sfe^\la$, $\la\in[0,1]$, via
$E_k^\la:=\la E_k+(1-\la)\id/3$, i.e.,
$$
E_0^\la=\begin{pmatrix}
a^\la_+ & 0 \\
0 & a^\la_-
\end{pmatrix}=
\frac13\begin{pmatrix}
1+\la & 0 \\
0 & 1-\la
\end{pmatrix},  \quad
E_1^\la=
\frac16\begin{pmatrix}
2-\la&-\sqrt3\la \\
-\sqrt3\la&2+\la
\end{pmatrix},  \quad
E_2^\la=
\frac16\begin{pmatrix}
2-\la&\sqrt3\la \\
\sqrt3\la&2+\la
\end{pmatrix}
$$
where $a^\la_\pm:=(1\pm\la)/3$. One can also write
$$
E_0^\la=\frac13\id+\frac\la 3\sigma_3,\quad E_1^\la=\frac13\id-\frac{\sqrt3\la}6\sigma_1-\frac\la6\sigma_3,\quad E_2^\la=\frac13\id+\frac{\sqrt3\la}6\sigma_1-\frac\la6\sigma_3.
$$
If $\la\ne 1$, $\sfe^\la$ is of rank 2 so that its minimal covariant Naimark dilation consists of 
the dilation space $\C^2\times\C^2\times\C^2\cong \C^2\oplus\C^2\oplus\C^2$, with the basis
$\{|k\pm\>\}$ where $|0\pm\>:=(|\pm\>,0,0)$, $|1\pm\>:=(0,|\pm\>,0)$, $|2\pm\>=(0,0,|\pm\>)$,
the PVM $P_k:=\kb{k+}{k+}+\kb{k-}{k-}$, $k\in\Z_3$, the isometry 
$$J_\la=\sum_{k=0}^2\sqrt{a^\la_+}\kb{k+}{+}U_k^*+\sqrt{a^\la_-}\kb{k-}{-}U_k^*,$$ 
and the unitary representation $$
\Z_3\ni k\mapsto V_k:=\sum_{\ell=0}^2\kb{b_k(\ell)+}{{\ell{+}}}+\kb{b_k(\ell)-}{{\ell{-}}}.
$$
Indeed,  clearly $P_k=V_k P_0 V_k^*$ is covariant, $J_\la U_k=V_k J_\la$, and $E^\la_k=J_\la^*P_k J_\la$.

Suppose then that $\sfb=(B_j)$ is a qubit POVM which is jointly measurable with $\sfe^\la$. For any\footnote{Of course, the joint POVM of $(\sfe^\la,\sff)$ needs not be unique.} joint POVM $\mathsf N=(N_{kj})$ (i.e., $\sum_{j}N_{kj}=E^\la_k$ and $\sum_{k}N_{kj}=B_j$) there exist three unique qubit POVMs $\tilde\sfa^{(k)}=\big(\tilde A^{(k)}_j\big)$, $k\in\Z_p$, such that $N_{kj}=J_\la^*\tilde A^{(k)}_j J_\la$; here $\tilde \sfa^{(k)}$ operates in the subspace spanned by $\{|k+\>,|k-\>\}$.
By defining matrices 
$$
A_j^{(k)}:=\begin{pmatrix}
\< k{+|}\tilde A_j^{(k)}|k+\> & \< k{+|}\tilde A_j^{(k)}|k-\>\\
\< k{-|}\tilde A_j^{(k)}|k+\> & \< k{-|}\tilde A_j^{(k)}|k-\>
\end{pmatrix},
$$
we get 
$$
N_{kj}=J_\la^*\tilde A^{(k)}_j J_\la=U_k \big(M^\la\star A^{(k)}_j\big)U_k^*
$$ 
where 
$$
M^\la:=\begin{pmatrix}
{a^\la_+} & \sqrt{a^\la_+a^\la_-} \\
\sqrt{a^\la_-a^\la_+} & {a^\la_-}
\end{pmatrix}
=
\frac13\begin{pmatrix}
1+\la & \sqrt{1-\la^2} \\
\sqrt{1-\la^2} & 1-\la
\end{pmatrix}\ge 0
$$
and $\star$ is the entrywise (Schur) product. 
One can solve the $A$--matrices (showing uniqueness):
$$
A^{(k)}_j=(U_k^*N_{kj} U_k\big)\star N^\la
$$
where
$$
N^\la:=
3\begin{pmatrix}
(1+\la)^{-1} & 1/\sqrt{1-\la^2} \\
1/\sqrt{1-\la^2} & (1-\la)^{-1}
\end{pmatrix}\ge 0.
$$
Note that there may be many $A$--matrices (i.e., many joint POVMs $\sfg$) giving the same marginal
$B_j= \big(M^\la\star A^{(0)}_j\big)+U_1 \big(M^\la\star A^{(k)}_j\big)U_1^*+U_2 \big(M^\la\star A^{(2)}_j\big)U_2^*$.

\begin{example}
Let $\sfb=\sfb^\eta$ above, where $\psi$ is a unit vector, $\eta\in[0,1]$, and
$$
B_0^\eta=\eta\frac23\kb\psi\psi+(1-\eta)\frac13\id,\qquad B_k^\eta=U_k B_0^\eta U_k^*,\quad k\in\Z_3.
$$
Since $\sfb^\eta$ is also covariant we may assume that $\mathsf N$ is covariant, i.e., satisfies 
$N_{k+\ell,j+\ell}=U_\ell N_{kj} U_\ell^*$ (note that 
$\tilde N_{kj}:=\frac13\sum_{s=0}^2 U_s^* N_{k+s,j+s} U_s$ gives the same marginals as $\mathsf N$ and is covariant).

For $\psi=|-\>$, we guess that 
$$
A^{(k)}_0=U_k^* A U_k, 
\qquad
A=\begin{pmatrix}
d & 0 \\
0 & e 
\end{pmatrix},
\qquad d,\,e\ge 0,\quad d+e=\frac23,
$$
could be a good choice for 
$N_{k0}=U_k \big(M^\la\star A^{(k)}_0\big)U_k^*$. This determines the rest of the effects by the covariance condition $N_{k+\ell,j+\ell}=U_\ell N_{kj} U_\ell^*$.
Let us solve parameter $d$ from
\begin{eqnarray*}
B_0^\eta&=&\frac13\begin{pmatrix}
1-\eta & 0 \\
0 & 1+\eta
\end{pmatrix}
=M^\la\star A+U_1 \big(M^\la\star U_1^* A U_1\big)U_1^*+U_2 \big(M^\la\star U_2^* A U_2\big)U_2^* \\
&=&\frac14\begin{pmatrix}
\left(3+\sqrt{1-\la^2}\right)d+\left(1-\sqrt{1-\la^2}\right)e & 0 \\
0 & \left(3+\sqrt{1-\la^2}\right)e+\left(1-\sqrt{1-\la^2}\right)d
\end{pmatrix}
\end{eqnarray*}
where $e=\frac23-d$. The solution is
$$
d=\frac{\la^2-2\eta\left(1-\sqrt{1-\la^2}\right)}{3 \la^2}\le\frac{\la^2}{3 \la^2}= \frac13
$$
so that automatically $e=\frac23-d\in[\frac13,\frac23]$ when $d\in[0,\frac13]$.
The condition $d\ge 0$ is equivalent to $\la^2-2\eta\left(1-\sqrt{1-\la^2}\right)\ge 0$ or
$$
\eta\le \frac{\la^2}{2\left(1-\sqrt{1-\la^2}\right)}=:f(\la)
$$
where $f(\la)$ runs from $1$ to $\frac12$ when $\la$ goes from $0$ to $1$, i.e., the POVMs are jointly measurable when $\eta\in[0,f(\la)]$.
If we assume that $\eta=\la$ then one must have $\la\le f(\la)$, i.e., $\la\le 4/5=0.8$. This, however, is not the optimal value, as can be seen analytically by using the joint measurement characterisation of Ref.~\cite{jae19}, which gives $\lambda\lesssim 0.866$. 
When $\eta\in[0,f(\la)]$ we can easily calculate the joint POVM $\mathsf N$. 

\end{example}

\section{Examples of jointly measurable continuous qubit POVMs}

Suppose that two qubit POVMs $\sfe$ and $\sfb$ (defined on $\sigma$-algebras $\hA\subseteq 2^\Omega$ and $\hB\subseteq 2^\Xi$) are jointly measurable with a joint POVM $\No$.
Now a Naimark dilation of $\sfe$ can be constructed as follows:
Let $\mu:\,\hA\to[0,1]$ be a probability measure such that $\sfe$ is absolutely continuous with respect to it. For example, $\mu(X)=\frac12\tr{\sfe(X)}$ is fine.
Then, by the Radon-Nikodym theorem, $\sfe$ has a qubit density $D$, i.e., one can write
$\sfe(X)=\int_X D(x)\d\mu(x),$ $x\in X,$
where each $D(x)$ is positive semidefinite $2\times 2$ matrix and $x\mapsto D(x)$ is $\mu$-measurable \cite{Pello3}. By using the spectral decomposition of $D(x)$, also $x\mapsto\sqrt{D(x)}$ is $\mu$-measurable and one can define an isometry $J$ from $\C^2$ into $L^2(\mu)\otimes\C^2$, where $L^2(\mu)$ is the Lebesgue (Hilbert) space,\footnote{Consisting of equivalence classes of $\mu$-square integrable complex functions on $\Omega$.} 
via 
$$
\left[J
\begin{pmatrix}
c_1 \\
c_2
\end{pmatrix}
\right](x):=\sqrt{D(x)}\begin{pmatrix}
c_1 \\
c_2
\end{pmatrix},\qquad (c_1,c_2)\in\C^2,\quad x\in\Omega.
$$
Now, for all $X\in\hA$, one gets 
$\sfe(X)=J^*\Po(X)J$ 
where $\Po$ is the canonical spectral measure defined by
$$
\left[
\Po(X)\begin{pmatrix}
\psi_1 \\
\psi_2
\end{pmatrix}
\right](x):=
\begin{pmatrix}
\CHI X(x)\psi_1(x) \\
\CHI X(x)\psi_2(x)
\end{pmatrix},\qquad \psi_1,\,\psi_2\in L^2(\mu),
$$
and $\CHI X$ is the characteristic function of the set $X\in\hA$. The obtained Naimark dilation is not necessarily minimal\footnote{The only drawback of non-minimality is that, for a given $\No$, the POVMs $\Fo_x$ are note necessarily unique.} (since the rank of $D(x)$ can be 0 or 1) but  we can still write, for all $X\in\hA$ and $Y\in\hB$,
$$
\No(X\times Y)=\int_X\sqrt{D(x)}\Fo_x(Y)\sqrt{D(x)}\d\mu(x),\qquad
\sfb(Y)=\int_\Omega\sqrt{D(x)}\Fo_x(Y)\sqrt{D(x)}\d\mu(x),
$$
where each $\Fo_x$ is a qubit POVM\footnote{Actually, one needs some reqularity conditions for the measurable spaces, e.g., they are standard Borel \cite{Pello7}.} on $\hB$ \cite{pellonpaa14,Pello7}.
In the next examples, $\sfe$ and $\sfb$ are given and we try to find $\No$ by assuming that each $\Fo_x$ is absolutely continuous with respect to the probability measure $\nu$ of $\sfb$ (e.g., $\nu(Y)=\frac12\tr{\sfb(Y)}$); now $\No$ is absolutely continuous with respect to the product measure $\mu\times\nu$ and has a qubit density with respect to it.
This method is easy to generalise for three (or more) jointly measurable POVMs as follows.

In the rest of this section, we study a single-mode optical field (i.e., a harmonic oscillator) in the case of a single photon. 
We use the position representation of the position (quadrature) operator $Q$ (with the spectral measure $\sfq$) where the number states $|n\rangle$ are represented as the Hermite--Gauss functions
$$
h_n(x) := \frac{1}{\sqrt{2^nn!\sqrt{\pi}}}H_n(x)e^{-\frac 12 x^2}
=\frac{(-1)^n}{\sqrt{2^nn!\sqrt{\pi}}} e^{\frac12x^2}\frac{\d^n e^{-x^2}}{\d x^n}
$$
where $H_n$ is a (normalised) Hermite polynomial. Hermite polynomials $H_n$ are given
by the recursion relation $H_0(x)=1$, $H_1(x)=2x$, and $H_{n+1}(x)= 2xH_n(x)-2nH_{n-1}(x)$ or by the Rodrigue's formula $H_n(x)=(-1)^n e^{x^2}\d^n e^{-x^2}/\d x^n$. 
The spectral measure of the rotated (or tilted) quadrature $Q_\theta$ is
$$
\sfq_\theta(X)=e^{i\theta a^* a}\sfq(X)e^{-i\theta a^* a}
=\sum_{n,m=0}^\infty e^{i(n-m)\theta}\int_X h_n(x)h_m(x)\d x \kb n m
$$
for all Borel sets $X\subseteq\R$; here $a$ is the lowering operator \cite{busch16}.
Note that the momentum operator $P=Q_{\pi/2}$.
The projection of $\sfq_\theta$ onto the single-photon subspace ${\rm span}\{ |0\rangle,|1\rangle\}$ is the (qubit) POVM
$$
 \int_X \begin{pmatrix} 1 & \sqrt2 x e^{-i\theta} \\  \sqrt2 x e^{i\theta} & 2x^2 \end{pmatrix} 
 \frac{e^{-x^2}\d x}{\sqrt\pi}
$$
whose noisy version is
\begin{eqnarray*}
\sfq_\theta^{\rm prono}(X)&:=&(1-\epsilon_\theta)\underbrace{
 \int_X \begin{pmatrix} 1 & \sqrt2 x e^{-i\theta} \\  \sqrt2 x e^{i\theta} & 2x^2 \end{pmatrix} 
 \frac{e^{-x^2}\d x}{\sqrt\pi}
}_{\text{noiseless projected $\sfq_\theta$}}+\epsilon_\theta\underbrace{\int_X\begin{pmatrix}1 & 0 \\ 0 & 1  \end{pmatrix} \frac{e^{-x^2}\d x}{\sqrt\pi}
}_{\text{the random noise}} \\
&=&
 \int_X \begin{pmatrix} 1 & (1-\epsilon_\theta)\sqrt2 x e^{-i\theta} \\  (1-\epsilon_\theta)\sqrt2 x e^{i\theta} & (1-\epsilon_\theta)2x^2+\epsilon_\theta \end{pmatrix} 
 \frac{e^{-x^2}\d x}{\sqrt\pi}
\end{eqnarray*}
where $\epsilon_\theta\in[0,1]$.

Similarly consider energy $H=\hbar\omega\Big(a^* a+\frac12\Big)$ whose spectral measure is 
$\{n\}\mapsto \No_n:=\kb n n$.
The projected noisy energy POVM has two non-zero effects:
\begin{eqnarray*}
\No^{\rm prono}_0&:=&(1-\epsilon)\underbrace{\begin{pmatrix}1 & 0 \\ 0 & 0  \end{pmatrix}}_{\text{vacuum effect}} + \epsilon \underbrace{\frac12 \begin{pmatrix}1 & 0 \\ 0 & 1 \end{pmatrix}}_{\text{noise effect}}
=\begin{pmatrix}1-\epsilon/2 & 0 \\ 0 & \epsilon/2  \end{pmatrix} \\
\No^{\rm prono}_1&:=&(1-\epsilon)\underbrace{\begin{pmatrix}0 & 0 \\ 0 & 1  \end{pmatrix}}_{\text{1-photon effect}} + \epsilon \underbrace{\frac12 \begin{pmatrix}1 & 0 \\ 0 & 1 \end{pmatrix}}_{\text{noise effect}}
=\begin{pmatrix} \epsilon/2 & 0 \\ 0 & 1-\epsilon/2  \end{pmatrix} 
\end{eqnarray*}
where $\epsilon\in[0,1]$. Next we study joint measurability of the triplet $\big(\No^{\rm prono},\sfq_0^{\rm prono},\sfq_\theta^{\rm prono}\big)$ for any $\theta\notin\{0,\pi\}$.

Using a result of \cite[Appendix B]{kiukas17} we get a joint normalised operator valued  measure (OVM) $\G$ whose effects are 
{\tiny
\begin{eqnarray*}
&&\G\big(\{0\}\times X\times Y\big)=\int_Y\int_X\frac{\d x\d y}\pi e^{-x^2-y^2}\times\\
&&\times\begin{pmatrix}1-\epsilon/2 & (1-\epsilon/2) \big[(1-\epsilon_0)\sqrt2x+(1-\epsilon_\theta)\sqrt2ye^{-i\theta}\big] \\ 
(1-\epsilon/2) \big[(1-\epsilon_0)\sqrt2x+(1-\epsilon_\theta)\sqrt2ye^{i\theta}\big]
 & \epsilon/2+ (1-\epsilon/2) \big[(1-\epsilon_0)2x^2+(1-\epsilon_\theta)2y^2+(1-\epsilon_0)(1-\epsilon_\theta)4xy\cos\theta+\epsilon_0+\epsilon_\theta-2\big]
\end{pmatrix},\\
&&\G\big(\{1\}\times X\times Y\big)=\int_Y\int_X\frac{\d x\d y}\pi e^{-x^2-y^2}\times\\
&&\times\begin{pmatrix}\epsilon/2 & \epsilon/2 \big[(1-\epsilon_0)\sqrt2x+(1-\epsilon_\theta)\sqrt2ye^{-i\theta}\big] \\ 
\epsilon/2 \big[(1-\epsilon_0)\sqrt2x+(1-\epsilon_\theta)\sqrt2ye^{i\theta}\big]
 & 1-\epsilon/2+ \epsilon/2 \big[(1-\epsilon_0)2x^2+(1-\epsilon_\theta)2y^2+(1-\epsilon_0)(1-\epsilon_\theta)4xy\cos\theta+\epsilon_0+\epsilon_\theta-2\big]
\end{pmatrix}.
\end{eqnarray*}
}

Since $\int_\R x e^{-x^2}\d x = 0$ and $\int_\R 2x^2 e^{-x^2}\d x/\sqrt\pi = 1$ one gets a joint OVM for $\sfq_0^{\rm prono}$ and $\sfq_\theta^{\rm prono}$: {\tiny
\begin{eqnarray*}
&&\G_1\big( X\times Y\big):=\G\big(\{0,1\}\times X\times Y\big)=\G\big(\{0\}\times X\times Y\big)+\G\big(\{1\}\times X\times Y\big)=\int_Y\int_X\frac{\d x\d y}\pi e^{-x^2-y^2}\times\\
&&\times\begin{pmatrix}1 & (1-\epsilon_0)\sqrt2x+(1-\epsilon_\theta)\sqrt2ye^{-i\theta} \\ 
(1-\epsilon_0)\sqrt2x+(1-\epsilon_\theta)\sqrt2ye^{i\theta}
 & (1-\epsilon_0)2x^2+(1-\epsilon_\theta)2y^2+(1-\epsilon_0)(1-\epsilon_\theta)4xy\cos\theta+\epsilon_0+\epsilon_\theta-1
\end{pmatrix},\\
&&\G_1\big( X\times \R\big)=\G\big(\{0,1\}\times X\times \R\big)=\int_X\frac{\d x}{\sqrt\pi} e^{-x^2}
\begin{pmatrix}1 & (1-\epsilon_0)\sqrt2x \\ 
(1-\epsilon_0)\sqrt2x
 & (1-\epsilon_0)2x^2+\epsilon_0
\end{pmatrix}=\sfq_0^{\rm prono}(X),\\
&&\G_1\big( X\times \R\big)=\G\big(\{0,1\}\times X\times \R\big)=\sfq_\theta^{\rm prono}(Y).
\end{eqnarray*}
}
Similarly, a joint OVM of $\No^{\rm prono}$ and $\sfq_\theta^{\rm prono}$
{\tiny
\begin{eqnarray*}
&&\G_2\big(\{0\}\times Y\big):=\G\big(\{0\}\times \R\times Y\big)=\int_Y\frac{\d y}{\sqrt\pi} e^{-y^2}\begin{pmatrix}1-\epsilon/2 & (1-\epsilon/2) (1-\epsilon_\theta)\sqrt2ye^{-i\theta} \\ 
(1-\epsilon/2) (1-\epsilon_\theta)\sqrt2ye^{i\theta}
 & \epsilon/2+ (1-\epsilon/2) \big[(1-\epsilon_\theta)2y^2+\epsilon_\theta-1\big]
\end{pmatrix},\\
&&\G_2\big(\{1\}\times Y\big):=\G\big(\{1\}\times \R\times Y\big)=\int_Y\frac{\d y}{\sqrt\pi} e^{-y^2}\begin{pmatrix}\epsilon/2 & \epsilon/2 (1-\epsilon_\theta)\sqrt2ye^{-i\theta} \\ 
\epsilon/2(1-\epsilon_\theta)\sqrt2ye^{i\theta}
 & (1-\epsilon/2)+ \epsilon/2 \big[(1-\epsilon_\theta)2y^2+\epsilon_\theta-1\big]
\end{pmatrix},\\
&&\G_2\big(\{0\}\times\R\big)=\G\big(\{0\}\times \R\times \R\big)=\begin{pmatrix}1-\epsilon/2 & 0\\ 
0
 & \epsilon/2
\end{pmatrix}=\No^{\rm prono}_\theta,\qquad \G_2\big(\{1\}\times\R\big)=
\G\big(\{1\}\times \R\times \R\big)=\No^{\rm prono}_1
\\
\end{eqnarray*}
}
Finally, for $\No^{\rm prono}$ and $\sfq_0^{\rm prono}$ we get
{\tiny
\begin{eqnarray*}
&&\G_3\big(\{0\}\times X\big):=\G\big(\{0\}\times X\times \R\big)=\int_X\frac{\d x}{\sqrt\pi} e^{-x^2}\begin{pmatrix}1-\epsilon/2 & (1-\epsilon/2) (1-\epsilon_0)\sqrt2x \\ 
(1-\epsilon/2) (1-\epsilon_0)\sqrt2x
 & \epsilon/2+ (1-\epsilon/2) \big[(1-\epsilon_0)2x^2+\epsilon_0-1\big]
\end{pmatrix},\\
&&\G_3\big(\{1\}\times X\big):=\G\big(\{1\}\times X\times \R\big)=\int_X\frac{\d x}{\sqrt\pi} e^{-x^2}\begin{pmatrix}\epsilon/2 & \epsilon/2 (1-\epsilon_0)\sqrt2x \\ 
\epsilon/2(1-\epsilon_0)\sqrt2x
 & (1-\epsilon/2)+ \epsilon/2 \big[(1-\epsilon_0)2x^2+\epsilon_0-1\big]
\end{pmatrix}.
\end{eqnarray*}
}
When OVMs $\G$, $\G_1$, $\G_2$, and $\G_3$ are positive, i.e., POVMs? 
Since these OVMs contain a matrix inside an integral, the matrix must be positive semidefinite for all values of $x$ and $y$. But the left upper corners of the matrices are always non-negative so it is enough to check that the determinants are non-negative:

For $\G\big(\{0\}\times X\times Y\big)$ and $\G\big(\{1\}\times X\times Y\big)$ the determinants are
\begin{eqnarray*}
&&\underbrace{\left(1-\frac{\epsilon}{2}\right)^2}_{>0}\left[
\frac{\epsilon}{2-\epsilon}+\epsilon_0+\epsilon_\theta-2
+\underbrace{\epsilon_0(1-\epsilon_0)2x^2
+\epsilon_\theta(1-\epsilon_\theta)2y^2}_{\ge 0}
\right]
\end{eqnarray*}
and
\begin{eqnarray*}
&&\underbrace{\left(\frac{\epsilon}{2}\right)^2}_{\ge0}\left[
\underbrace{\frac{2-\epsilon}{\epsilon}}_{\ge\frac{\epsilon}{2-\epsilon}}+\epsilon_0+\epsilon_\theta-2
+\underbrace{\epsilon_0(1-\epsilon_0)2x^2
+\epsilon_\theta(1-\epsilon_\theta)2y^2}_{\ge 0}
\right]
\end{eqnarray*}
which are non-negative (i.e., $\G$ is a POVM) exactly when
$$
\boxed{\frac{\epsilon}{2-\epsilon}+\epsilon_0+\epsilon_\theta-2\ge 0.}
$$
The determinant related to $\G_1$ is
$$
 \epsilon_0 + \epsilon_\theta -1 + (1 - \epsilon_0) \epsilon_0 2x^2 +  (1 - \epsilon_\theta) \epsilon_\theta 2y^2
 \ge \epsilon_0 + \epsilon_\theta -1 
$$
so that $\G_1$ is positive exactly when
$$
\boxed{ \epsilon_0 + \epsilon_\theta -1 \ge 0.}
$$
Since $\epsilon_0 + \epsilon_\theta -1\ge \frac{\epsilon}{2-\epsilon}+\epsilon_0+\epsilon_\theta-2$,\footnote{Where the equality holds iff $\epsilon=1$ (i.e., the effects $\No_{0,1}^{\rm prono}=\frac12\id$ which commute with everything).}
the positivity of $\G$ implies the positivity of $\G_1$ as expected. If $\epsilon<1$ and $\epsilon_0=\epsilon_\theta=\frac12$ then $\G_1$ is positive but $\G$ is not.

For $\G_2$ we get (from determinants) the positivity conditions
$$
\underbrace{\frac2{2-\epsilon}}_{\le2/\epsilon}+\epsilon_\theta-2+\underbrace{
(1-\epsilon_\theta)\epsilon_\theta 2 y^2}_{\ge 0}\ge 0,\qquad
\frac2{\epsilon}+\epsilon_\theta-2+
(1-\epsilon_\theta)\epsilon_\theta 2 y^2\ge 0
$$
showing that $\G_2$ is a POVM if and only if 
$$
\boxed{\frac2{2-\epsilon}+\epsilon_\theta-2\ge 0.}
$$
Note that, since $\frac2{2-\epsilon}+\epsilon_\theta-2\ge\frac{\epsilon}{2-\epsilon}+\epsilon_0+\epsilon_\theta-2$,\footnote{Where the equality holds iff $\epsilon_0=1$ (i.e., $\sfq_0^{\rm prono}$ is a trivial POVM which commutes with everything).} clearly 
$\G_2$ is positive if $\G$ is positive but the converse does not hold (e.g., if $\epsilon=\frac12$, $\epsilon_\theta=\frac23$, and $\epsilon_0<1$).
Similarly, $\G_3$ is a POVM exactly when 
$$
\boxed{\frac2{2-\epsilon}+\epsilon_0-2\ge 0.}
$$
\begin{example}
Suppose that $\epsilon=\epsilon_0=\epsilon_\theta$.
Now $\No^{\rm prono},$ $\sfq_0^{\rm prono},$ and $\sfq_\theta^{\rm prono}$ are jointy measurable ($\G$) if
$
-2\epsilon^2+7\epsilon-4\ge 0,
$
i.e., when
$$
\boxed{
\epsilon\ge\frac14\left(7 - \sqrt{17}\right)\approx0.72.
}
$$
In addition, $\sfq_0^{\rm prono}$ and $\sfq_\theta^{\rm prono}$ are compatible ($\G_1$)
if 
$$
\epsilon\ge\frac12,
$$
and $\No^{\rm prono}$ and $\sfq_\theta^{\rm prono}$ (or $\sfq_0^{\rm prono}$) are jointly measurable if 
$
2\ge (2-\epsilon)^2,
$
i.e., when 
$$
\epsilon\ge2-\sqrt2\approx0.59.
$$
\hfill$\Box$\end{example}

Note that $\G$ (based on \cite{kiukas17}) is not the best possible joint measurement since, e.g., its marginal $\G_1$ gives a joint measurement for $\sfq_0^{\rm prono}$ and $\sfq_\theta^{\rm prono}$ for any $\theta$
only when
$\epsilon_0 + \epsilon_\theta  \ge 1$ but we know that, if $\theta=0$ or $\theta=\pi$, then
$\sfq_0^{\rm prono}$ and $\sfq_\theta^{\rm prono}$ are jointly measurable {\it for all}
$\epsilon_0,\,\epsilon_\theta\in[0,1]$. Let us try to find a better joint measurement (i.e., we will modify $\G$).

Suppose for simplicity that $\epsilon=\epsilon_0=\epsilon_\theta$ like in the preceding example.
Define an OVM via
{\tiny
\begin{eqnarray*}
&&\G'\big(\{0\}\times X\times Y\big)=\int_Y\int_X\frac{\d x\d y}\pi e^{-x^2-y^2}
\begin{pmatrix}1-\epsilon/2 & f(\epsilon ) \big[\sqrt2x+\sqrt2ye^{-i\theta}\big] \\ 
f(\epsilon )\big[\sqrt2x+\sqrt2ye^{i\theta}\big]
 &  f(\epsilon )\big[2x^2+2y^2+4xy\cos\theta\big]+g(\epsilon)
\end{pmatrix},    \\
&&\G'\big(\{1\}\times X\times Y\big)=\int_Y\int_X\frac{\d x\d y}\pi e^{-x^2-y^2}
\begin{pmatrix}\epsilon/2 & h(\epsilon ) \big[\sqrt2x+\sqrt2ye^{-i\theta}\big] \\ 
h(\epsilon )\big[\sqrt2x+\sqrt2ye^{i\theta}\big]
 &  h(\epsilon )\big[2x^2+2y^2+4xy\cos\theta\big]+i(\epsilon)
\end{pmatrix}   
\end{eqnarray*}
}
where $f$, $g$, $h$, and $i$ are unknown real functions. Since
{\tiny
\begin{eqnarray*}
&&\G'\big(\{0,1\}\times X\times Y\big)=\int_Y\int_X\frac{\d x\d y}\pi e^{-x^2-y^2}
\begin{pmatrix}1 & [f(\epsilon )+h(\epsilon)] \big[\sqrt2x+\sqrt2ye^{-i\theta}\big] \\ 
[f(\epsilon )+h(\epsilon)]\big[\sqrt2x+\sqrt2ye^{i\theta}\big]
 &  \big[f(\epsilon )+h(\epsilon )\big]\big[2x^2+2y^2+4xy\cos\theta\big]+g(\epsilon)+i(\epsilon)
\end{pmatrix}
\end{eqnarray*}
}
we must have 
$$
f(\epsilon)+h(\epsilon)=1-\epsilon,\qquad
g(\epsilon)+i(\epsilon)=2\epsilon-1,\qquad
2f(\epsilon )+g(\epsilon)=\epsilon/2,
$$
that is,
$$
g(\epsilon)=\epsilon/2-2f(\epsilon),\qquad
h(\epsilon)=1-\epsilon-f(\epsilon),\qquad
i(\epsilon)=3\epsilon/2-1+2f(\epsilon)
$$
where $f(\epsilon)$ is free. The determinants of the matrices inside $\G'\big(\{0\}\times X\times Y\big)$ and $\G'\big(\{1\}\times X\times Y\big)$ are
\begin{eqnarray*}
&&\big[(1-\epsilon/2)f(\epsilon)-f(\epsilon)^2\big]\big|\sqrt2x+\sqrt2ye^{-i\theta}\big|^2+(1-\epsilon/2)g(\epsilon),
\\
&&\big[(\epsilon/2)h(\epsilon)-h(\epsilon)^2\big]\big|\sqrt2x+\sqrt2ye^{-i\theta}\big|^2+(\epsilon/2)i(\epsilon)
\end{eqnarray*}
which must be non-negative for all $x,\,y\in\R$. Especially, by putting $x=y=0$ (or looking the lower right corners) we get the necessary conditions
$g(\epsilon)\ge 0$ and $i(\epsilon)\ge 0$, that is,
$
\epsilon/4\ge 
f(\epsilon)\ge 1/2-3\epsilon/4
$
which can hold if and only if $\epsilon/4\ge 1/2-3\epsilon/4$, i.e., $\epsilon\ge 1/2$.\footnote{Let us see what happens if we put $\epsilon=1/2$ implying
$$
f(1/2)=1/8,\qquad g(1/2)=0,\qquad h(1/2)=3/8,\qquad i(1/2)=0,
$$
so the second determinant reduce to $-(3/64)\big|\sqrt2x+\sqrt2ye^{-i\theta}\big|^2$ which is negative if, e.g., $(x,y)=(1,0)$.}

The next conditions are {\it sufficient} for $\G'$ being positive:
\begin{eqnarray*}
&&\epsilon> 1/2,\qquad
\epsilon/4\ge 
f(\epsilon)\ge 1/2-3\epsilon/4,\\
&&(1-\epsilon/2)f(\epsilon)-f(\epsilon)^2\ge 0
\quad\Longleftrightarrow\quad 0\le f(\epsilon)\le 1-\epsilon/2,
\\
&&(\epsilon/2)h(\epsilon)-h(\epsilon)^2\ge 0
\quad\Longleftrightarrow\quad 0\le h(\epsilon)\le \epsilon/2
\quad\Longleftrightarrow\quad 0\le 1-\epsilon-f(\epsilon)\le \epsilon/2
\end{eqnarray*}
which reduce to
$
\epsilon> 1/2$ and $\min\{\epsilon/4,1-\epsilon/2,1-\epsilon\}\ge
f(\epsilon)\ge\max\{1/2-3\epsilon/4,0,1-3\epsilon/2\}
$
or, equivalently,
$$
\begin{cases}
1-3\epsilon/2\le f(\epsilon)\le \epsilon/4, & 1/2<\epsilon\le 2/3, \\
0 \le  f(\epsilon)\le \epsilon/4, & 2/3<\epsilon\le 4/5, \\
0 \le  f(\epsilon)\le 1-\epsilon, & 4/5<\epsilon\le 1.
\end{cases}
$$
To conclude, to find some $f(\epsilon)$ (and a joint POVM $\G'$) one must have 
$1-3\epsilon/2\le \epsilon/4$, i.e., 
$$
\boxed{
\epsilon\ge 4/7\approx 0.57...
}
$$
so we have a better limit. Now $f(4/7)=1/7$,  $g(4/7)=0$, $h(4/7)=2/7$, $i(4/7)=1/7$, and the joint POVM is
{\tiny
\begin{eqnarray*}
&&\G'\big(\{0\}\times X\times Y\big)=\int_Y\int_X\frac{\d x\d y}\pi e^{-x^2-y^2}
\begin{pmatrix}5/7 & (1/7) \big[\sqrt2x+\sqrt2ye^{-i\theta}\big] \\ 
(1/7)\big[\sqrt2x+\sqrt2ye^{i\theta}\big]
 &  (1/7)\big[2x^2+2y^2+4xy\cos\theta\big]
\end{pmatrix},    \\
&&\G'\big(\{1\}\times X\times Y\big)=\int_Y\int_X\frac{\d x\d y}\pi e^{-x^2-y^2}
\begin{pmatrix}2/7 & (2/7) \big[\sqrt2x+\sqrt2ye^{-i\theta}\big] \\ 
(2/7)\big[\sqrt2x+\sqrt2ye^{i\theta}\big]
 &  (2/7)\big[2x^2+2y^2+4xy\cos\theta\big]+1/7
\end{pmatrix}.   
\end{eqnarray*}
}
By using similar methods, we find a better limit $\epsilon=\frac12$ and a joint POVM for the pair $\No^{\rm prono}$ and $\sfq_\theta^{\rm prono}$: 
{\tiny
\begin{eqnarray*}
&&\G'_2\big(\{0\}\times Y\big)=\int_Y\frac{\d y}{\sqrt\pi} e^{-y^2}\begin{pmatrix}3/4 & (1/4)\sqrt2ye^{-i\theta} \\ 
(1/4) \sqrt2ye^{i\theta}
 & (1/4) 2y^2
\end{pmatrix},\\
&&\G'_2\big(\{1\}\times Y\big)=\int_Y\frac{\d y}{\sqrt\pi} e^{-y^2}\begin{pmatrix}1/4 & (1/4)\sqrt2ye^{-i\theta} \\ 
(1/4)\sqrt2ye^{i\theta}
 & (1/4)2y^2+1/2
\end{pmatrix}.
\end{eqnarray*}
}
For $\sfq_0^{\rm prono}$ and $\sfq_\theta^{\rm prono}$ this method gives the same limit $\epsilon=1/2$ as before with a joint POVM
{\tiny
\begin{eqnarray*}
&&\G'_1\big( X\times Y\big)=\int_Y\int_X\frac{\d x\d y}\pi e^{-x^2-y^2}\frac12\begin{pmatrix}2 & \sqrt2x+\sqrt2ye^{-i\theta} \\ 
\sqrt2x+\sqrt2ye^{i\theta}
 & 2x^2+2y^2+4xy\cos\theta
 \end{pmatrix}.
\end{eqnarray*}
}
Finally, we study compatibility of the projected energy (or number) and the canonical phase.

Let $\epsilon\in[0,1]$ and denote $\epsilon^\perp:=1-\epsilon$.
Define the noisy projected canonical phase \cite{busch16}: for all Borel $X\subseteq[0,2\pi)$,
$$
\Phi(X):=\epsilon^\perp\underbrace{\int_X\begin{pmatrix}1 & e^{-i\theta} \\ e^{i\theta} & 1  \end{pmatrix}\frac{\d\theta}{2\pi}}_{\text{the canonical phase}}+\epsilon\underbrace{\int_X\begin{pmatrix}1 & 0 \\ 0 & 1  \end{pmatrix}\frac{\d\theta}{2\pi}}_{\text{the random noise}}
=\int_X\begin{pmatrix}1 & \epsilon^\perp e^{-i\theta} \\ \epsilon^\perp e^{i\theta} & 1  \end{pmatrix}\frac{\d\theta}{2\pi}\ge 0
$$
for which $\Phi\big([0,2\pi)\big)=\id$. Let $c$ and $d$ be complex numbers in the unit disc, i.e., $|c|\le1$ and $|d|\le 1$.
Define a  POVM via
\begin{eqnarray*}
\Mo\big(\{0\}\times X\big)&:=&\int_X\begin{pmatrix}1-\epsilon/2 & c\sqrt{\epsilon/2}\sqrt{1-\epsilon/2} e^{-i\theta} \\ \overline c\sqrt{\epsilon/2}\sqrt{1-\epsilon/2} e^{i\theta} & \epsilon/2  \end{pmatrix}\frac{\d\theta}{2\pi}\ge 0, \\
\Mo\big(\{1\}\times X\big)&:=&\int_X\begin{pmatrix}\epsilon/2 & d\sqrt{\epsilon/2}\sqrt{1-\epsilon/2} e^{-i\theta} \\ \overline d\sqrt{\epsilon/2}\sqrt{1-\epsilon/2} e^{i\theta} & 1-\epsilon/2  \end{pmatrix}\frac{\d\theta}{2\pi}\ge 0. 
\end{eqnarray*}
[For positivity note that, e.g., the determinant of the above matrix is $(\epsilon/2)(1-\epsilon/2)(1-|d^2|)\ge 0$.]
Now $\Mo\big(\{0\}\times[0,2\pi)\big)=\No^{\rm prono}_0$ and $\Mo\big(\{1\}\times[0,2\pi)\big)=\No^{\rm prono}_1$.
To get a joint measurement of $\No^{\rm prono}$ and $\Phi$ we must have also $\Mo\big(\{0\}\times X\big)+\Mo\big(\{1\}\times X\big)=\Phi(X)$ for all $X$, i.e., for all $\theta$,
{\tiny
$$
\begin{pmatrix}1-\epsilon/2 & c\sqrt{\epsilon/2}\sqrt{1-\epsilon/2} e^{-i\theta} \\ \overline c\sqrt{\epsilon/2}\sqrt{1-\epsilon/2} e^{i\theta} & \epsilon/2  \end{pmatrix}+
\begin{pmatrix}\epsilon/2 & d\sqrt{\epsilon/2}\sqrt{1-\epsilon/2} e^{-i\theta} \\ \overline d\sqrt{\epsilon/2}\sqrt{1-\epsilon/2} e^{i\theta} & 1-\epsilon/2  \end{pmatrix} = \begin{pmatrix}1 & \epsilon^\perp e^{-i\theta} \\ \epsilon^\perp e^{i\theta} & 1  \end{pmatrix},
$$
}
i.e.,
$
(c+d)\sqrt{\epsilon/2}\sqrt{1-\epsilon/2}=\epsilon^\perp=1-\epsilon
$ or
$$
c+d=f(\epsilon):=\frac{1-\epsilon}{\sqrt{\epsilon/2}\sqrt{1-\epsilon/2}}.
$$
Since $|c+d|\le|c|+|d|\le 2$, one has to have $f(\epsilon)\le 2$ [i.e., then $c$ and $d$ and $\Mo$ exist; one can choose, e.g., $c=d=f(\epsilon)/2$].
So to get the smallest possible $\epsilon$ (see the next figure) we must solve equation $f(\epsilon)=2$. The solution is
$\epsilon=\epsilon_{\rm min}:=1-1/\sqrt{2}\approx 0.292893$.

To conclude, if $\epsilon\ge\epsilon_{\rm min}$ then $\No^{\rm prono}$ and $\Phi$ are jointly measurable.

\begin{figure}[htbp]
\begin{center}
\includegraphics[width=0.8\textwidth]{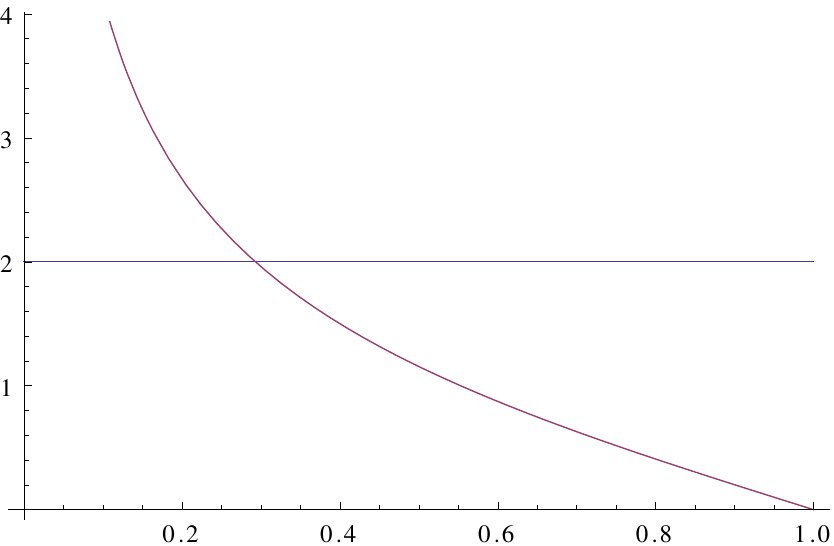} 
\caption{The function $\epsilon\mapsto f(\epsilon)$ [the red curve].}
\label{kuva}
\end{center}
\end{figure}

\section{Conclusions}

We have analysed a method for characterising all measurements that are jointly measurable with a given measurement. The technique maps the problem of deciding joint measurability into the problem of finding suitable block-diagonal POVMs in a minimal Naimark dilation space of one of the involved POVMs. We have demonstrated the use of the technique with heuristic ans\"atze. Whereas some of these lead to the optimal noise tolerance, such as in the case of the celebrated Busch criterion \cite{busch86}, we have shown that in other scenarios, such as symmetric trinary qubit measurements, an optimal ansatz may be harder to find. We have further presented a full closed-form characterisation of all qubit effects that are jointly measurable with a given qubit effect, and extended our analysis to scenarios involving pairs and triplets of continuous qubit measurements.

\section*{Acknowledgements}

The authors are grateful for the fruitful discussions with M\'at\'e Farkas and Felix Huber. R.U.\ is thankful for the support from the Swiss National Science Foundation (Ambizione PZ00P2-202179).

\bibliographystyle{unsrt}
\bibliography{References.bib}

\begin{thebibliography}{10}

\bibitem{busch16}
Paul {Busch}, Pekka {Lahti}, Juha-Pekka {Pellonp\"a\"a}, and Kari {Ylinen}.
\newblock {\em Quantum Measurement (Theoretical and Mathematical Physics)}.
\newblock Springer, 2016.

\bibitem{buschrmp2014}
Paul Busch, Pekka Lahti, and Reinhard~F. Werner.
\newblock Colloquium: Quantum root-mean-square error and measurement
  uncertainty relations.
\newblock {\em Rev. Mod. Phys.}, 86:1261--1281, 2014.

\bibitem{coles17}
Patrick~J.\ Coles, Mario Berta, Marco Tomamichel, and Stephanie Wehner.
\newblock Entropic uncertainty relations and their applications.
\newblock {\em Rev. Mod. Phys.}, 89:015002, 2017.

\bibitem{Gisin2002Review}
Nicolas Gisin, Gr\'egoire Ribordy, Wolfgang Tittel, and Hugo Zbinden.
\newblock Quantum cryptography.
\newblock {\em Rev. Mod. Phys.}, 74:145--195, 2002.

\bibitem{Degen2017}
C.L. Degen, F.~Reinhard, and P.~Cappellaro.
\newblock Quantum sensing.
\newblock {\em Reviews of Modern Physics}, 89(3), 2017.

\bibitem{RandomnessReview}
Miguel Herrero-Collantes and Juan~Carlos Garcia-Escartin.
\newblock Quantum random number generators.
\newblock {\em Rev. Mod. Phys.}, 89:015004, 2017.

\bibitem{Guryanova20}
Yelena Guryanova, Nicolai Friis, and Marcus Huber.
\newblock Ideal projective measurements have infinite resource costs.
\newblock {\em Quantum}, 4:222, 2020.

\bibitem{oszmaniec19}
Micha{\l} Oszmaniec and Tanmoy Biswas.
\newblock Operational relevance of resource theories of quantum measurements.
\newblock {\em Quantum}, 3:133, 2019.

\bibitem{uola19b}
Roope Uola, Tristan Kraft, Jiangwei Shang, Xiao-Dong Yu, and Otfried G\"uhne.
\newblock Quantifying quantum resources with conic programming.
\newblock {\em Phys. Rev. Lett.}, 122:130404, 2019.

\bibitem{Pekka03}
Pekka Lahti.
\newblock Coexistence and joint measurability in quantum mechanics.
\newblock {\em Int. J. Theor. Phys.}, 42:893, 2003.

\bibitem{heinosaari10}
Teiko Heinosaari and Michael M.\ Wolf.
\newblock Non-disturbing quantum measurements.
\newblock {\em J. Math. Phys.}, 51:092201, 2010.

\bibitem{heinosaari16a}
Teiko Heinosaari.
\newblock Simultaneous measurement of two quantum observables: compatibility,
  broadcasting, and in-between.
\newblock {\em Phys. Rev. A}, 93:042118, 2016.

\bibitem{oszmaniec17}
Micha{\l} Oszmaniec, Leonardo Guerini, Peter Wittek, and Antonio Ac\'in.
\newblock Simulating positive-operator-valued measures with projective
  measurements.
\newblock {\em Phys. Rev. Lett.}, 119:190501, 2017.

\bibitem{Ioannou22}
Marie Ioannou, Pavel Sekatski, S\'ebastien Designolle, Benjamin D.~M. Jones,
  Roope Uola, and Nicolas Brunner.
\newblock Simulability of high-dimensional quantum measurements. arxiv:
  2202.12980.

\bibitem{Cope22}
Thomas Cope and Roope Uola.
\newblock Quantifying the high-dimensionality of quantum devices.
  arxiv:2207.05722.

\bibitem{wolf09}
Michael M.\ Wolf, David Perez-Garcia, and Carlos Fernandez.
\newblock Measurements incompatible in quantum theory cannot be measured
  jointly in any other local theory.
\newblock {\em Phys. Rev. Lett.}, 103:230402, 2009.

\bibitem{quintino14}
Marco T\'ulio {Quintino}, Tam\'as {V\'ertesi}, and Nicolas {Brunner}.
\newblock Joint measurability, Einstein-Podolsky-Rosen steering, and bell
  nonlocality.
\newblock {\em Phys. Rev. Lett.}, 113:160402, 2014.

\bibitem{uola14}
Roope {Uola}, Tobias {Moroder}, and Otfried {G\"uhne}.
\newblock Joint measurability of generalized measurements implies classicality.
\newblock {\em Phys. Rev. Lett.}, 113:160403, 2014.

\bibitem{uola15}
Roope {Uola}, Costantino {Budroni}, Otfried {G\"uhne}, and Juha-Pekka {Pellonp\"a\"a}.
\newblock One-to-one mapping between steering and joint measurability problems.
\newblock {\em Phys. Rev. Lett.}, 115:230402, 2015.

\bibitem{kiukas17}
Jukka {Kiukas}, Costantino {Budroni}, Roope {Uola}, and Juha-Pekka {Pellonp\"a\"a}.
\newblock Continuous-variable steering and incompatibility via state-channel
  duality.
\newblock {\em Phys. Rev. A}, 96:042331, 2017.

\bibitem{tavakoli19}
Armin Tavakoli and Roope Uola.
\newblock Measurement incompatibility and steering are necessary and sufficient
  for operational contextuality.
\newblock {\em Physical Review Research}, 2(1), 2020.

\bibitem{skrzypczyk19}
Paul Skrzypczyk, Ivan \v{S}upi\'c, and Daniel Cavalcanti.
\newblock All sets of incompatible measurements give an advantage in quantum
  state discrimination.
\newblock {\em Phys. Rev. Lett.}, 122:130403, 2019.

\bibitem{carmeli19a}
Claudio {Carmeli}, Teiko {Heinosaari}, and Alessandro {Toigo}.
\newblock Quantum incompatibility witnesses.
\newblock {\em Phys. Rev. Lett.}, 122:130402, 2019.

\bibitem{guerini19}
Leonardo Guerini, Marco~T{\'u}lio Quintino, and Leandro Aolita.
\newblock Distributed sampling, quantum communication witnesses, and
  measurement incompatibility.
\newblock {\em Phys. Rev. A}, 100:042308, 2019.

\bibitem{Beyer22}
Konstantin Beyer, Roope Uola, Kimmo Luoma, and Walter~T. Strunz.
\newblock Joint measurability in nonequilibrium quantum thermodynamics.
\newblock {\em Physical Review E}, 106(2), 2022.

\bibitem{Zhou16}
Fei Zhou, Leilei Yan, Shijie Gong, Zhihao Ma, Jiuzhou He, Taiping Xiong, Liang
  Chen, Wanli Yang, Mang Feng, and Vlatko Vedral.
\newblock Verifying Heisenberg's error-disturbance relation using a single
  trapped ion.
\newblock {\em Science Advances}, 2(10):e1600578, 2016.

\bibitem{Designolle21}
S\'ebastien Designolle, Vatshal Srivastav, Roope Uola, Natalia~Herrera
  Valencia, Will McCutcheon, Mehul Malik, and Nicolas Brunner.
\newblock Genuine high-dimensional quantum steering.
\newblock {\em Phys. Rev. Lett.}, 126:200404, 2021.

\bibitem{Hammad2020}
Hammad Anwer, Sadiq Muhammad, Walid Cherifi, Nikolai Miklin, Armin Tavakoli,
  and Mohamed Bourennane.
\newblock Experimental characterization of unsharp qubit observables and
  sequential measurement incompatibility via quantum random access codes.
\newblock {\em Phys. Rev. Lett.}, 125:080403, 2020.

\bibitem{Smirne2022}
Andrea Smirne, Simone Cialdi, Daniele Cipriani, Claudio Carmeli, Alessandro
  Toigo, and Bassano Vacchini.
\newblock Experimentally determining the incompatibility of two qubit
  measurements.
\newblock {\em Quantum Sci. Technol.}, 7:025016, 2022.

\bibitem{JMreview}
Otfried G\"uhne, Erkka Haapasalo, Tristan Kraft, Juha-Pekka Pellonp\"a\"a, and
  Roope Uola.
\newblock Incompatible measurements in quantum information science.
  arxiv:2112.06784.

\bibitem{pellonpaa14}
Juha-Pekka Pellonp\"a\"a.
\newblock On coexistence and joint measurability of rank-1 quantum observables.
\newblock {\em J. Phys. A}, 47(5):052002, 2014.

\bibitem{haapasalo15}
Erkka Haapasalo, Juha-Pekka Pellonp\"a\"a, and Roope Uola.
\newblock Compatibility properties of extreme quantum observables.
\newblock {\em Lett. Math. Phys.}, 105:661--673, 2015.

\bibitem{Pello7}
Erkka Haapasalo and Juha-Pekka Pellonp\"a\"a.
\newblock Optimal quantum observables.
\newblock {\em J. Math. Phys.}, 58(12):122104, 2017.

\bibitem{JMNaimark}
P.~Kruszy\'nski and W.~M. de~Muynck.
\newblock Compatibility of observables represented by positive operator valued
  measures.
\newblock {\em Journal of Mathematical Physics}, 28(8):1761--1763, 1987.

\bibitem{Beneduci14}
Roberto Beneduci.
\newblock Joint measurability through Naimark's dilation theorem.
\newblock {\em Reports on Mathematical Physics}, 79(2):197--214, 2017.

\bibitem{Mitra20}
Arindam Mitra, Sibasish Ghosh, and Prabha Mandayam.
\newblock Characterizing incompatibility of quantum measurements via their
  naimark extensions. arxiv:2011.11364.

\bibitem{busch86}
Paul Busch.
\newblock Unsharp reality and joint measurements for spin observables.
\newblock {\em Phys. Rev. D}, 33:2253, 1986.

\bibitem{jae19}
Jeongwoo Jae, Kyunghyun Baek, Junghee Ryu, and Jinhyoung Lee.
\newblock Necessary and sufficient condition for joint measurability.
\newblock {\em Phys. Rev. A}, 100:032113, 2019.

\bibitem{Uola16}
Roope {Uola}, Kimmo {Luoma}, Tobias {Moroder}, and Teiko {Heinosaari}.
\newblock Adaptive strategy for joint measurements.
\newblock {\em Phys. Rev. A}, 94:022109, 2016.

\bibitem{Heinosaari13b}
Teiko Heinosaari.
\newblock A simple sufficient condition for the coexistence of quantum effects.
\newblock {\em Journal of Physics A: Mathematical and Theoretical},
  46(15):152002, 2013.

\bibitem{designolle19b}
S{\'e}bastien Designolle, M{\'a}t{\'e} Farkas, and Jedrzej Kaniewski.
\newblock Incompatibility robustness of quantum measurements: a unified
  framework.
\newblock {\em New J. Phys.}, 21:113053, 2019.

\bibitem{Kiukas22}
Jukka Kiukas, Daniel McNulty, and Juha-Pekka Pellonp\"a\"a.
\newblock Amount of quantum coherence needed for measurement incompatibility.
\newblock {\em Physical Review A}, 105(1), 2022.

\bibitem{Chiribella10}
Giulio Chiribella, Giacomo~Mauro D'Ariano, and Dirk Schlingemann.
\newblock Barycentric decomposition of quantum measurements in finite
  dimensions.
\newblock {\em Journal of Mathematical Physics}, 51(2):022111, 2010.

\bibitem{Pello3}
Tuomas Hyt{\"o}nen, Juha-Pekka Pellonp\"a\"a, and Kari Ylinen.
\newblock Positive sesquilinear form measures and generalized eigenvalue
  expansions.
\newblock {\em J. Math. Anal. Appl.}, 336:1287--1304, 2007.

\end{thebibliography}

\newpage
\end{document}